\documentclass[12pt]{iopart}
\usepackage{iopams}  
\usepackage[dvipdfmx]{graphicx}
\usepackage{psfrag}
\usepackage{xcolor}

\newcommand{\Tinv}[1]{\tan^{-1}\left({#1}\right)}

\newcommand{\lb}{\mathcal{L}}
\newcommand{\rb}{\mathcal{R}}

\def\lb{\mathcal{L}}
\def\rb{\mathcal{R}}
\def\S{\mathcal{S}}
\def\B{\mathcal{B}}
\def\Z{\mathcal{Z}}
\def\P{\mathcal{P}}
\def\D{\mathcal{D}}
\def\A{\mathcal{A}}
\def\W{\mathcal{W}}
\def\G{\Gamma}

\usepackage{ulem}

\begin{document}

\title{Particle transport in open polygonal billiards: a scattering map}
\author{Jordan Orchard$^{1}$, Federico Frascoli$^{1}$, Lamberto Rondoni$^{2,3}$, Carlos Mej\'ia-Monasterio$^{4,5}$}
\address{$^1$ Department of Mathematics, School of Science, Computing and Engineering Technologies, Swinburne University of Technology, H38, PO Box 218, Hawthorn, Victoria 3122, Australia}
\address{$^2$ Dipartimento di Scienze Matematiche, Politecnico di Torino, Corso Duca degli Abruzzi 24, 10129 Torino, Italy}
\address{$^3$ INFN, Sezione di Torino, Via P. Giuria 1, 10125 Torino, Italy}
\address{$^4$
School of Agricultural, Food and Biosystems Engineering,
Technical University of Madrid, Av. Puerta de Hierro 2, 28040 Madrid, Spain}
\address{$^{5}$ Grupo Interdisciplinar de Sistemas Complejos (GISC), Spain}

\ead{\mailto{jorchard@swin.edu.au},\mailto{ffrascoli@swin.edu.au},\mailto{lamberto.rondoni@polito.it}, \mailto{carlos.mejia@upm.es}}
\vspace{10pt}

\begin{indented}
\item[]\today
\end{indented}

\begin{abstract}
  Polygonal billiards exhibit a rich and complex dynamical behavior. In recent years polygonal billiards have attracted great attention due to their application in the understanding of anomalous transport, but also at the fundamental level, due to its connections with diverse fields in mathematics. We explore this complexity and its consequences on the properties of particle transport in infinitely long channels made of the repetitions of an elementary open polygonal cell. Borrowing ideas from the Zemlyakov-Katok construction, we construct an interval exchange transformation classified by the singular directions of the discontinuities of the billiard flow over the translation surface associated to the elementary cell. From this, we derive an exact expression of a scattering map of the cell connecting the outgoing flow of trajectories with the unconstrained incoming flow. The scattering map is defined over a partition of the coordinate space, characterized by different families of trajectories. Furthermore, we obtain an analytical expression for the average speed of propagation of ballistic modes, describing with high accuracy the speed of propagation of ballistic fronts appearing in the tails of the distribution of the particle displacement. The symbolic hierarchy of the trajectories forming these ballistic fronts is also discussed.
\end{abstract}
\vspace{2pc}

\noindent{Submitted to: CHAOS: An Interdisciplinary Journal of Nonlinear Science.}
%

\section{Introduction}
\label{sec:intro}
Mathematical billiards provide simple yet non-trivial examples of dynamical systems \cite{bunimovich2000,tabachnikov2005a,chernov2006}.  Assuming the billiard as a planar bounded domain $\Omega$, the billiard dynamics is generated by the free motion of a point massless particle subject to elastic reflections in the boundary $\delta\Omega$. This means that the particle moves with constant speed, along a straight line specularly reflected in the boundary.

Billiards have widely been used to study the foundations of the statistical mechanics \cite{balint2021}, ergodicity and mixing \cite{sinai1970,bunimovich1979,bunimovich1992,young1998,sinai2007,simanyi2013,baladi2018}, to understand transport phenomena \cite{vanbeyeren1972,dorfman1997,gaspard1998,dettmann2000,artuso2000,alonso2002,armstead2003,sanders2006,zarfaty2019}, and to model nonequilibrium systems \cite{mejia2001,larralde2003,eckmann2006,eckmann2006b,collet2009,collet2009b,gaspard2009}, to mention a few.  While billiards are a crude idealization of real systems, they shed light on \emph{e.g.} the behavior of gases in the Knudsen limit in which the interactions between the particles and the walls are comparable when not greater, than particle-particle interactions \cite{jepps2003,jepps2006,jepps2008, comets2010,su2012,chumley2015}.

A particular class of billiards is that of polygonal billiards with a boundary $\delta \Omega$ consisting on linear segments. Polygonal billiards are piecewise continuous, the discontinuities arising from the corners at the vertices of the polygons.  In polygonal billiards, corners split nearby trajectories into different families that separate in time slower than exponential, yielding zero Lyapunov exponents characteristic of integrable billiards. However, the discontinuous billiard flow due to the corners break the otherwise smooth and regular tori of the phase space of integrable systems, and conical singularities appear.

The mathematical challenge of determining the dynamics of polygonal billiards is due to their intermediate, neither integrable nor chaotic, behavior. Their complexity has raised great interest in the past decades \cite{galperin1983,gutkin1986,vorobets1997, smillie1999,zorich2006,bobok2012,fraczek2014,attarchi2020}, particularly because of their connections with recent mathematical advances, such as translation surfaces, Riemannian manifolds and Teichm\"uller flows on moduli spaces \cite{zemlyakov1975,kontsevich2003,wright2015,eskin2018,conze2012,hubert2011,delecroix2014,matheus2018}.

In this paper, we focus on a specific class of periodic polygonal billiard infinite channels, consisting of an identical elementary polygonal cell connected through its openings. The polygonal cell consists of two identical boundaries, made of two edges forming an angle $\alpha$, separated by a distance $d$. In recent years, we have considered these polygonal channels to study the transport of particles \cite{vollmer2021,orchard2021}. Channels with not necessarily parallel upper and lower boundaries, have also been investigated recently \cite{sanders2006,jepps2006,jepps2008,alonso2002}. Particle transport in polygonal channels depends strongly and erratically on the details of their geometry. With parallel boundaries, transport exhibits strong anomalous diffusion \cite{castiglione1999}, with a superdiffusive mean square displacement and ballistic higher order moments of the distribution of the particle displacement $P(x,t)$ \cite{vollmer2021,orchard2021}. The ballistic scaling is evident in the tails of $P(x,t)$ in which several ballistic fronts, with different speed of propagation, can be observed. With non-parallel boundaries, diffusion and even subdiffusion has been observed \cite{sanders2006,jepps2006,jepps2008,alonso2002}.

Here we make an improvement for the evaluation of the billiard dynamics by deriving an exact scattering map of the elementary cell. The scattering map is set up by unfolding the polygonal billiard into a translation surface, following the Zemlyakov-Katok construction \cite{zemlyakov1975}, and then derived as an interval exchange transformation describing the evolution of billiard trajectories \cite{keane1975,veech1978}. Translation surfaces provide a natural setting for studying the dynamics of polygonal billiards, as we will see. \cite{smillie1999,hubert2011,delecroix2014,zahradova2022,zahradova2023}. 

The scattering map is defined over a partition of the coordinate-space that we study in detail. Each set in the partition is characterized by a specific family of trajectories sharing the same sequence of edges the trajectory is reflected from before leaving the cell. These sequences are finite, except for some sets containing bouncing trajectories of arbitrary length. This is done for one polygonal cell with finite horizon and one with infinite horizon.
Following this, we revisit particle transport along an infinitely long channel and focus on the ballistic fronts discussed previously in Ref.~\cite{orchard2021}. Finally, we derive an analytical expression for the speed of propagation of the so-called sub-leading ballistic front in $P(x,t)$ and analyze the families of trajectories contributing to this ballistic front. 

The paper is organized as follows. In Sec.~\ref{sec:model} we define the polygonal billiards that we study and sketch our program to obtain the scattering map. In Sec.~\ref{sec:finite} we construct the interval exchange transform for a polygonal cell with finite horizon, and derive the scattering map. The derivation is repeated in Sec.~\ref{sec:infinite} for the cell with infinite horizon. In Sec.~\ref{sec:dx} we discuss particle transport in an polygonal channel. In Sec.~\ref{sec:concl} we summarize our results.

\section{Open polygonal billiards}
\label{sec:model}

\begin{figure}[!ht]
    \centering
    \includegraphics[width=0.7\textwidth]{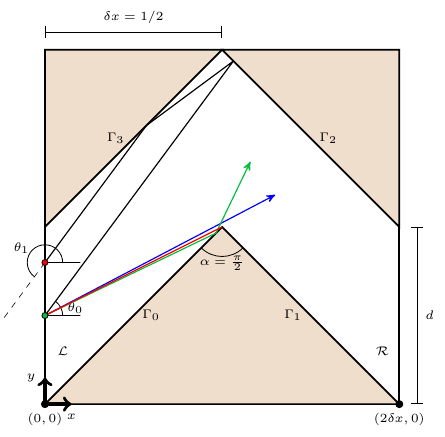}
    \caption{Geometry of the elementary cell characterized by the opening angle $\alpha$ and width $d$, shown here for $\alpha=1/2$ and $d = 1/2$.  The edges of the boundary are labeled $\{ \G_i \}_{i=0}^3$, and the `left' and `right' openings are labeled $\{\lb, \rb \}$. The black polyline shows a scattered trajectory with itinerary $\lb \G_2 \G_3 \lb$. The green and blue lines show the separation of trajectories passing near a vertex. The singular direction of the vertex is shown as a red line. The length of the cell is fixed to $\delta x = 1/2$. The height of the bottom polygon is $\delta y = \delta x \cot \frac{\alpha}{2} = \delta x$. }
     \label{fig:schem}
\end{figure}

We consider billiard dynamics occurring in a polygonal channel with open ends. The billiard channel is made of identical copies of an elementary polygonal cell $\P$, with boundary $\delta\P$ formed by four straight sides and two openings as shown in Fig.~\ref{fig:schem}. The sides form identical upper and lower boundaries characterized by the opening angle $\alpha$. The open left and right ends of the cell, denoted $\lb$ and $\rb$ respectively, connect to the previous and following cells. Setting the semi-width of the cell to $2\delta x = 1$, the geometry is fully determined by $\alpha\in(0,\pi)$ and the separation $d$ between the upper and lower boundaries.

Inside the billiard, massless point particles fly freely with constant speed until they reach the boundary $\delta\P$, after which it reflects elastically if the segment of $\delta\P$ hit is the upper or lower boundaries, or crosses to an adjacent cell if it hits an opening. For $d > \delta x \cot(\alpha/2)$ the horizon is said to be infinite, meaning that a particle trajectory can cross the cell connecting $\lb$ and $\rb$ without hitting the cell boundaries. Conversely, for $d \le \delta x \cot(\alpha/2)$ the horizon is said to be finite, meaning that any trajectory entering the cell from $\lb$ or $\rb$ will experience a positive number of collisions with $\delta \P$ before exiting.

Trajectories inside the cell form families characterized by specific sequences of reflections from the different segments of the cell boundaries denoted as $\G_i$ with $ i=0,1,2,3$. Thus, billiard dynamics takes place inside the polygon $\mathcal{C} = \cup_{i=0}^3 \G_i$, and shifts periodically to neighbouring cells over the openings $\lb$ and $\rb$.
Figure \ref{fig:schem} shows the elementary cell for $\alpha=\pi/2$ and finite horizon $d=1/2$.\footnote{In the literature this is also referred to as critical horizon.}

Billiard dynamics in polygons have the numerical advantage that trajectories are exactly solved simply as the intersection of two linear segments, yielding a set of possible collisions well ordered in time without the necessity of event driven techniques. Therefore, trajectories are obtained with high numerical precision limited only by machine's logical precision. Nevertheless, infinite precision is in principle, required for close to singular trajectories.

Singular trajectories are those that reach one of the polygon's vertices, after which the dynamics stop. They are singular discontinuities of the billiard flow, and just as unstable periodic orbits determine the topology of the skeleton of heteroclinic orbits in hyperbolic billiards, singular trajectories determine the dynamical topology of polygonal billiards, separating nearby trajectories as they hit the boundary on opposite sides of a vertex, into families with different itineraries.

Figure~\ref{fig:schem} shows a singular trajectory (red line) emanating from $\lb$ at an angle $\theta_0$, and reaching the vertex at coordinates $(1/2,1/2)$. The two neighbouring trajectories (green and blue lines), separate into different itineraries: $\lb\G_0\G_2\lb$ for the green trajectory and $\lb\G_2 \G_1\rb$ for the blue trajectory.

To be specific, we define the openings $\lb$ and $\rb$ of the billiard cell in terms of generalized Birkhoff coordinates, the height along the opening $h \in \D \equiv (0, d)$ and direction of the trajectory $\theta \in \A \equiv \left[0, 2\pi\right)$. Identification of $\lb$ with $\rb$ through periodic boundary conditions make their $x$-coordinate irrelevant, which we demote in favour of a shift map $\sigma$ that updates the index corresponding to a particles location in the channel. 

Furthermore, let $\B = \D \times \A$ denote the coordinate space. The scattering map is defined as the function $\S_{d,\alpha}: \B \rightarrow \B$, where the dependence of the map on the cell geometry through $\alpha$ and $d$, has been made explicit. Clearly, $\S_{d,\alpha}$ is a permutation that rearranges the points in $\B$. In scattering, we need to distinguish incoming and outgoing trajectories. Namely, incoming trajectories are those passing over $\lb$, $\theta \in \A^+\equiv [-\pi/2,\pi/2)$, while outgoing trajectories are those passing over $\rb$, $\theta \in \A^-\equiv [\pi/2,3\pi/2)$, where $\A=\A^+\cup\A^-$.

Due to the symmetry of the billiard cell, $\S_{d,\alpha}$ is also symmetric with respect to $\theta=\pi/2$. In what follows, we redefine the scattering map $\widehat{\S}_{d,\alpha}: \B^+ \rightarrow \B$ as
\begin{equation}\label{eq:complete}
    \widehat{\mathcal{S}}_{d,\alpha}(h, \theta) \equiv \left\{
  \begin{array}{ll}
      \mathcal{S}(h, \theta)		      &\mathrm{if} \ \theta \in \mathcal{A}^+ \\	\\
    \mathcal{S}(h, \pi - \theta)  &\mathrm{if} \ \theta \in \mathcal{A}^- 	
  \end{array}\right. \ ,
\end{equation}
where $\B^+= \D \times \A^+$.

The billiard dynamics is solved by a finite number of iterations of the scattering map together with the application of periodic boundary conditions and the shift $\sigma$. Inside the elementary cell trajectories are labeled by the sequence of collisions with the boundaries $\{ \G_i\}_{i=0}^3$. The finiteness of the return time of the trajectory to one of the openings is a straightforward consequence of the geometry of the elementary cell, though in general, the Poincar\'e recurrence theorem applies \cite{feres2004, tabachnikov2005}.

In principle, the procedure described above can be applied to any geometry. Here, we will focus on the polygonal billiard with $\alpha=\pi/2$ and derive the scattering map for a cell with finite horizon, $d=1/2$ in Sec.~\ref{sec:finite}, and for a cell with infinite horizon $d=1$ in Sec.~\ref{sec:infinite}.

\section{The scattering map for a cell with finite horizon}
\label{sec:finite}

In this section we derive the scattering map $\widehat{\mathcal{S}}_{1/2,\pi/2}$ for the elementary cell with finite horizon $d=1/2$ and $\alpha=\pi/2$.

The derivation establishes an interval exchange transformation \cite{keane1975,veech1978}, where the intervals represent different regions of the billiard $\P$, that are accessible to families of trajectories sharing a similar sequence of collisions with the edges $\{\G_i\}$. The scattering map describes how the collisions permute these intervals.

We start by partitioning the domain $\B^+$ into a collection of disjoint open subsets $\B_j^+ = \mathcal{D}_j \times \mathcal{A}^+_j$, $\B^+= \bigcup_{j}\B_j^+$. The partitions are determined by the skeleton of singular trajectories. As we mentioned before, singular trajectories lead to discontinuities in the billiard flow, separating trajectories lying at each of its sides. For instance, consider the green and blue trajectory segments of Fig.~\ref{fig:schem}, lying at each side of the singular trajectory (in red). At successive times, their sequence of collisions will be different.  The crucial observation is that between any two singular trajectories, a family of regular trajectories, namely trajectories that never reach a vertex, exist.\footnote{Tangential or `grazing' trajectories, whereby a trajectory is tangent to the billiard boundary, may also constitute singular trajectories. However, in polygonal billiards we do not need to take special care of grazing trajectories as they are by definition, singular \cite{chernov2006}.}  Therefore, the sets $\B_j^+$ classify regular trajectories while their boundaries $\bigcup_{j} \partial \B_j^+$ constitute a set of zero measure corresponding to the singular trajectories.

To classify  different families of regular trajectories, we compare the symbolic sequence of collisions the trajectory suffers during its passage through one elementary cell.

For definiteness, we set initial incoming trajectories entering the cell from the left, so that their initial conditions are in $\B^+$.  All such sequences, or itineraries, are of the form $\lb \W \lb$ or $\lb \W \rb$, where $\W$ is a `word' comprised of the `letters' $\{ \Gamma_i \}_{i=0}^3$.  We use this symbolic dynamics of trajectories to set the taxonomy of different histories determining the partition $\{\B_j^+\}$ . From the point of view of transport, it is clear that histories of the form $\lb \mathcal{W} \rb$ contribute to the transmission of particles from left to right and $\lb \mathcal{W} \lb$, to reflection. The combinatorial properties of the different transmitting and reflecting histories fully determine the transport properties of the billiard.

To identify the partition $\{\B^+_j\}$, we consider a minimal covering in terms of the directions $\xi_m : \mathcal{D} \mapsto \mathcal{A}^+$ defining, for a given initial coordinate $h$, the initial directions corresponding to singular trajectories. We enumerate the singular directions such that $-\pi/2 \le \xi_m(h) \le \xi_{m+1} \le \pi/2$. Let
\begin{equation} \label{eq:beta}
  \rho_m(h) = \left\lbrace \theta :  \xi_m(h) \le \theta \le \xi_{m+1}(h) \right\rbrace, \ \mathrm{and} \quad  \beta_m = \mathcal{D} \times \rho_m \ .
\end{equation}
For a given initial coordinate $h$, the partition $\{\beta_m\}$ contains all trajectories that are not singular, and as such $\bigcup_{m}\beta_m \subset \B^+$ contains all non singular trajectories.

At this point one should be cautious as some $\beta_m$ may still contain singular trajectories. As we shall discuss in the next sections, this happens when one or more parametrisable singular directions subdivide further into new nested singular directions thus, subdividing $\beta_m$ into different regions. Notwithstanding this, defining a minimal covering as above is useful, since $\{\B^+_j\}$ coincides with the partition $\{\beta_m\}$ when no nested singular trajectories exist. If that is not the case then some $\beta_m$ will be subjected to further subdivisions. As we shall see, sets corresponding to the latter situation will be typically bounded by more than two families of singular directions.

In the remainder of the section we will explicitly derive the partition $\{\B^+_j\}$ and the scattering map $\S(h,\theta)$. It will we useful to divide the analysis depending on the segment of $\delta\P$ the incoming trajectories hit first, namely of the first symbol of $\mathcal{W}$. In the case of finite horizon this can be $\G_0, \G_2$ or $\G_3$.

\subsection{A first collision with the boundary $\Gamma_0$}
\label{sec:g0}

Let us consider the set of incoming trajectories that having initial conditions in $\B^+$ and which collide first with the boundary segment $\G_0$. To study the evolution of the billiard flow it is customary to use the Zemlyakov-Katok construction \cite{zemlyakov1975}. When a billiard trajectory hits a segment of the billiard boundary, instead of reflecting the direction of the trajectory, we can choose to reflect the billiard cell in that segment and let the billiard trajectory carry on straight. 
Following this construction, the billiard cell forms a surface, called a translation surface, on which the billiard trajectories follow rays (see \cite{gutkin1986}). 

For our purposes, we need only to unfold $\P$ until the ray $y(x) = x\tan \theta_0 + h_0$ intersects either $\lb$ or $\rb$ of the unfolded cell for the first time.  Denoting $P$ as the intersection point, the ray connecting $(h_0,\theta_0)$ and $P$ completely specifies the scattering trajectory. The latter can be recovered by folding back the translation surface, \emph{e.g.}, by reading $\W$ backwards. For this subset of incoming trajectories we find that the partition comprises eight regions separated by singular directions $\xi_0,\xi_1, \xi_2, ..., \xi_8$, with fixed $\xi_0=-\pi/2$.

\begin{figure}[!t]
    \centering
    \includegraphics[width=0.85\textwidth]{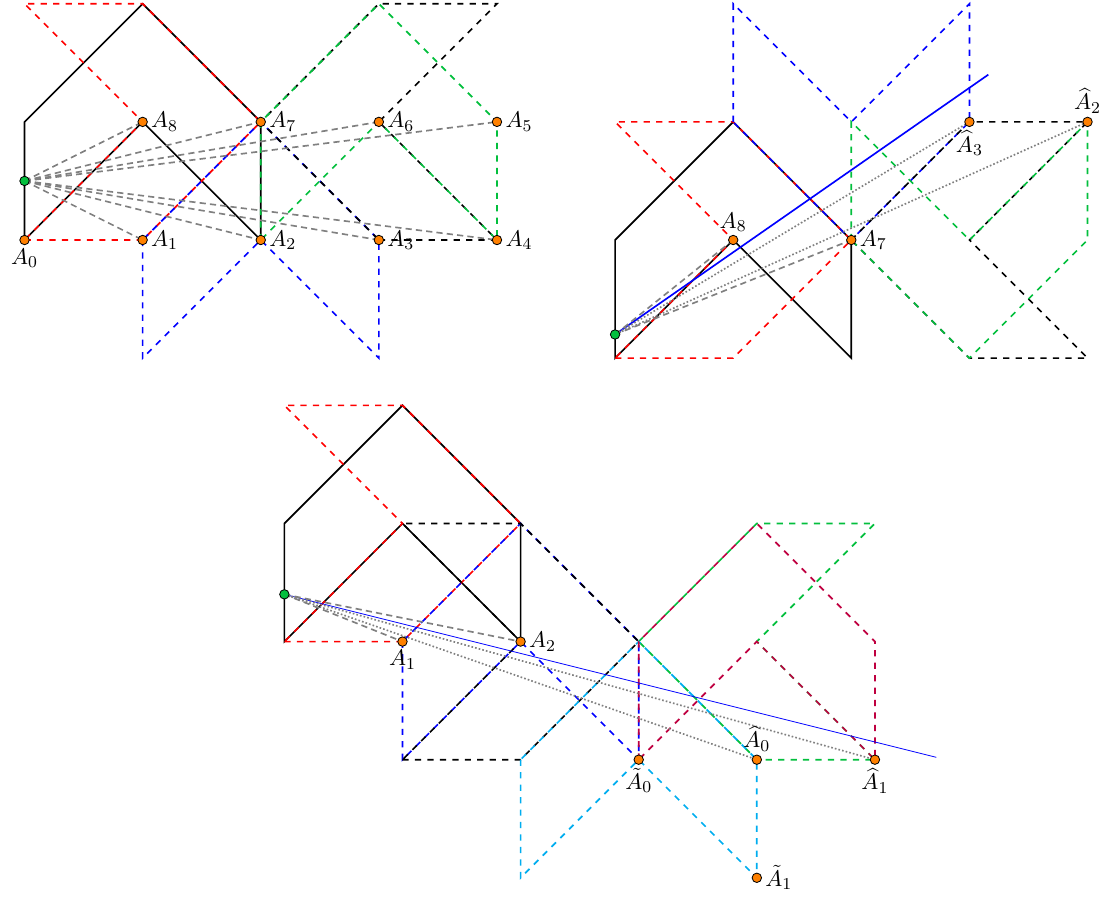}
    \caption{Translation surface of the elementary cell for incoming rays that first hit the edge $\G_0$. The points $\{A_i\}$ indicated the accessible vertices to the incoming flow. Top left: Unfolding for the horizontal rays $y = h \in \mathcal{D}$. The dashed gray lines depict the singular directions of the accessible vertices $\xi_m(h) = \angle \ h A_m$. Bottom: Splitting of the interval $\beta_1$ into three additional intervals as indicated by the vertices $\widehat{A}_{0,1}$.  Bouncing trajectories appear for rays passing between the points $\widetilde{A}_n= (1 + (n+1)/2, -(n+1)d)$, with $n=\{0,1,\ldots\}$, here shown $\widetilde{A}_0$ and $\widetilde{A}_1$. Top right: Splitting of the interval $\beta_7$ into three additional intervals as indicated by the vertices $\widehat{A}_{2,3}$.}
    \label{fig:g0unfold}
\end{figure}

First considering horizontal rays $y = h$, we find the translation surface shown in the top left panel of Fig.~\ref{fig:g0unfold}. The points $A_0,A_1, ..., A_8$ (orange circles), correspond to all vertices accessible to trajectories that experience a first collision with $\Gamma_0$. For example, the dashed gray line connecting the green dot and $A_5$ corresponds to a trajectory that terminates at the upper rightmost vertex of $\P$ following four regular collisions. And similarly for the remaining dashed gray lines which indicate singular directions for the trajectories starting with coordinate $h_0$ (green circle).
The coordinates of the vertices in the $(x,y)$ plane are
\begin{equation}
  A_m = \left\{
    \begin{array}{ll}
      \left(\frac{m}{2}, 0\right) & \mathrm{for} \ \ 1 \le m \le 4 \ , \\
      \\
      \left(\frac{9-m}{2}, d\right) & \mathrm{for} \ \ 4 < m \le 8 \ .
    \end{array}
  \right.
\end{equation}
The singular directions are readily obtained as the angle between the $x$-axis and the line connecting the initial point $(0,h)$ and $A_m$. Note that as required, $\xi_m \le \xi_{m+1}$ holds, so that the partition $\beta_m$ of Eq.~\ref{eq:beta} is well defined.

We now study the families of trajectories with initial conditions in each of the subsets $\beta_m$, and verify if $\beta_m$ does only contains regular trajectories, in which case $\beta_m = \B^+_m$.
As an example, consider the set $\beta_0$ which is the region bounded by the lines $\xi_0 = -\pi/2$ and $\xi_1(h) = -\tan^{-1} \frac{h}{d}$. From Fig.~\ref{fig:g0unfold} it is clear that any incoming trajectory in $\beta_m$ intersects the line segment $\overline{A_0 A_1}$ at the point $P = (h \cot \theta, 0)$.
Since $\overline{A_0 A_1}$ is the reflection of $\lb$ about $\G_0$, all incoming trajectories in $\beta_0$ are regular trajectories exiting the elementary cell through $\lb$ without experiencing further collisions.

Furthermore, reflecting $P$ about $\G_0$ gives the point $P' = (0, -h \cot \theta)$ which are the coordinates of the outgoing trajectory. Therefore, for $(h, \theta) \in \beta_0 = \B^+_0$ the scattering map is $\S_{1/2.\pi/2}(h, \theta) = (-h\cot \theta, \frac{\pi}{2} - \theta)$.

In the remainder of the article we will frequently omit the calculation of the points $P$ and $P'$ in each specific case, leaving the final result to be self-contained in the expression for $\S$. This includes further considerations of the sets $\beta_2, \beta_3,..., \beta_6$ which can be shown, through the above reasoning, to represent only regular trajectories and for the sake of convenience, we relabel the partition as $\B^+_{i+2} = \beta_i$ for $i = 2,3,...,6$.

For $\beta_1$ and $\beta_7$, we find that not all rays $y(x)$ pass through the same opening of the unfolded polygon. This automatically means that there exist singular trajectories not accounted by the partition generated by the points $A_m$.

Let us first examine the case of $\beta_1$.  From Fig.~\ref{fig:g0unfold} (bottom panel) we find additional singular directions $\widehat{\xi}_0(h)$, $\widehat{\xi}_1(h)$, corresponding to the angles of the lines passing through the points $\widehat{A}_{0}, \widehat{A}_1$ (dashed gray lines). This leads to a subdivision of $\beta_1$ into three regions corresponding to 
\begin{eqnarray} \label{eq:subbeta1}
  i) & \frac{1}{6}<h< d \ , \quad  \xi_1 < \theta < \widehat{\xi}_0 \ , \\
  ii) & \frac{1}{8}<h< d \ , \quad \widehat{\xi}_0 < \theta < \widehat{\xi}_1 \ , \\
  iii) & \ 0 <h<\frac{1}{3} \ , \quad \widehat{\xi}_1 < \theta < \xi_2 \ . 
\end{eqnarray}
The subdivisions of the coordinate $h$ are found by solving $\widehat{\xi}_{0,1}(h) = \xi_{1,2}(h)$ for $h$ with the corresponding indices.

Inspecting the regions $(ii)$ and $(iii)$, we realize that these are not fully contained in $\beta_1$. To fix this we further define $\A_2^+ = \{\theta : h \in (1/8,d) \ \mathrm{and} \ \widehat{\xi}_0 < \theta < \widehat{\xi}_1\} \cap \rho_1$ and $\A_3^+ = \{\theta : h \in (0,1/3)\ \mathrm{and} \ \widehat{\xi}_1 < \theta < \xi_2\} \cap \rho_1$. 

Finally, for region $(i)$ in Eq.~\ref{eq:subbeta1}, we define $\mathcal{A}_1^+ = \{\theta : h \in (1/6,d) \ \mathrm{and} \ \xi_1 < \theta < \widehat{\xi}_0\}$ and note that within $\beta_1$ there exist trajectories that bounce between the parallel walls $\Gamma_0$ and $\Gamma_3$ an arbitrary number of times, resulting in words $\{ \Gamma_0 \Gamma_3 \Gamma_0 \Gamma_3, \Gamma_0 \Gamma_3 \Gamma_0 \Gamma_3 \Gamma_0 \Gamma_3, \ldots \}$.  To deal with these bouncing trajectories we define the points $\widetilde{A}_n = (1 + (n+1)/2, -(n+1)d)$ with $n = 1,2,\ldots$ corresponding to the nested singular directions separating each periodic continuation of the above word. 

The points $\widehat{A}_{0,1}$ and $\widetilde{A}_{0,1}$ are shown in the bottom panel of Fig.~\ref{fig:g0unfold} where it can be seen that rays $y(x)$ passing between $\widetilde{A}_0$ and $\widehat{A}_0$ correspond to words $\Gamma_0 \Gamma_3 \Gamma_0 \Gamma_3$ while $\widetilde{A}_0$ and $\widetilde{A}_1$ determine words $\Gamma_0 \Gamma_3 \Gamma_0 \Gamma_3 \Gamma_0 \Gamma_3$ and so on, for each pair of points $\widetilde{A}_n$ and $\widetilde{A}_{n+1}$. This leads to yet another countable collection of singular directions $\widetilde{\xi}_n$ determined by the points $\widetilde{A}_n$, with $\widetilde{\xi}_0 = \widehat{\xi}_0$. These nested singular directions further subdivide the region $(i)$ in Eq.~\ref{eq:subbeta1} into an infinite cascade of regions that shrink as $(h,\theta) \to (d, -\pi/4)$. As before, we can solve $\widetilde{\xi}_n(h)=\xi_1(h)$ for $h$ to find the corresponding intervals of the coordinate $h$ that we denote as $\widetilde{\mathcal{D}}_n = \left(\frac{n}{2(n+1)}, d \right)$, valid for $n \ge 1$.  For $n = 0$ we have $\widetilde{\mathcal{D}}_0 = \mathcal{D}_1 = \left(\frac{1}{6},d\right)$. We denote this cascade of subregions of $\beta_1$ as $\beta_1^{(n)} = \widetilde{\mathcal{D}}_n \times \widetilde{\rho}_n$ where $\widetilde{\rho}_n = \{ \theta : h \in \widetilde{\mathcal{D}}_n \ \mathrm{and} \ \widetilde{\xi}_n(h) < \theta < \widetilde{\xi}_{n+1}(h) \} \cap \mathcal{A}_1^+$ and $\beta_1 = \bigcup_{n\ge0}\beta_1^{(n)} = \mathcal{D}_1 \times \mathcal{A}_1^+$.

In contrast with the previous regions, the unfolded image of bouncing trajectories do not have a unique intersection point $P$ with the openings, because each of the words found within $\mathcal{B}_1$ determines a different unfolding.  To find $P$ here, we proceed as follows: we intersect the ray $y(x)$ with a step function whose horizontal and vertical levels are $y=b_n = -nd$ and $x=a_n = 1 + nd$, respectively. This may lead to multiple intersections, of which the correct one is that with a smaller distance from the initial coordinates $(0, h)$ of the ray. This yields $y(a_n) \ge b_n$ that can be solved to find
\begin{equation} \label{eq:n}
  n \ge \frac{-h - \tan\theta}{d(1 + \tan \theta)} \ .
\end{equation}
Then defining the integer function
\begin{equation} \label{eq:iotag0}
  \iota(h,\theta) = \left\lceil \frac{-h - \tan\theta}{d(1 + \tan \theta)} \right\rceil
\end{equation}
where $\left\lceil x \right\rceil$ is the least integer greater than or equal to $x$, we obtain the intersection point of the ray with either $\lb$ or $\rb$ to be $P = (1+\iota d, y(1 + \iota d))$, and folding the translation surface back to the elementary cell the coordinates of the outgoing trajectory are $P' = (2\delta x, h + \frac{1}{2}\left[ \iota(1 + \tan \theta) + 3 \tan\theta + 1 \right])$. Note that $\iota(h,\theta) = n + 1$ when $(h,\theta) \in \beta_1^{(n)}$ for $n\ge0$.

As a final remark of the region $\beta_1$, we notice that the word $\G_0\G_3$, a natural candidate for this family of trajectories, is absent. It turns out that $\Gamma_0 \Gamma_3$ trajectories are instead associated with the region $\beta_4$ and the above expression $P'$ with $\iota = 0$ is indeed valid for this case. Thus, we think of $\beta_1 \cup \beta_4$ as a unified region characterizing the bouncing trajectories $\{ \Gamma_0 \Gamma_3 \Gamma_0 \Gamma_3, \Gamma_0 \Gamma_3 \Gamma_0 \Gamma_3 \Gamma_0 \Gamma_3, \ldots \} \cup \{ \G_0 \G_3 \}$, even if in practice, we will assign a different index to each region. 

We now turn to the last region, $\beta_7$. From the top right panel of Fig.~\ref{fig:g0unfold} we find a situation similar to $\beta_1$, in that $\beta_7$ subdivides into three regions due to the presence of two nested singular directions $\widehat{\xi}_{2}(h)$ and $\widehat{\xi}_{3}(h)$, connecting the initial coordinates of the ray $(0,h)$ with the points $\widehat{A}_2$ and $\widehat{A}_3$ respectively (dotted gray lines).

Proceeding as before, we solve $\widehat{\xi}_{2,3} = \xi_8(h)$ for $h$, and identify three further regions on which $\beta_7$ is subdivided: $\mathcal{D}_9 = \mathcal{D}$, $\mathcal{D}_{10} = \left(0, \frac{1}{3}\right)$, $\mathcal{D}_{11} = \left( 0, \frac{1}{4} \right)$ and, the angular domains
\begin{eqnarray}
  \mathcal{A}_9^+ & = \left\{ \theta : h \in \mathcal{D}_9 \ \mathrm{and} \ \xi_7 < \theta < \widehat{\xi}_2 \right\} \cap \rho_7 \ , \\
  \mathcal{A}_{10}^+ & = \left\{ \theta : h \in \mathcal{D}_{10} \ \mathrm{and} \ \widehat{\xi}_2 < \theta < \widehat{\xi}_3 \right\} \cap \rho_7 \ , \\
  \mathcal{A}_{11}^+ & = \left\{ \theta : h \in \mathcal{D}_{11} \ \mathrm{and} \ \widehat{\xi}_3 < \theta < \xi_8 \right\} \ .
\end{eqnarray}

Summarizing, we have analyzed trajectories whose initial conditions belong to the set $\mathcal{D} \times \{ \theta: \ \xi_0 < \theta < \xi_8\}$, corresponding to particles that have their first collision on the edge $\G_0$. In total, eleven distinct types of trajectories, grouped by their symbolic itineraries were identified (see Table~\ref{tab:iting0}) and shown to coincide with the partition regions $\beta_j, j = 0, 1, ..., 10$. Furthermore, we found that the region, $\beta_1 \cup \beta_4$ was characterized by the existence of a family of bouncing trajectories, having arbitrarily long word lengths.

Having identified the partition regions $\{\beta_m\}$, the explicit calculation of the scattering map $\mathcal{S}(h,\theta)$ is straightforward: for a ray $y(x) = x\tan\theta + h$ with $(h, \theta) \in \beta_j$ we look for the point of intersection between a vertical or horizontal line segment in the translation surface corresponding to the image of the unfolded openings $\lb$ or $\rb$. From this intersection point one can reverse the order and fold the translation surface back to the elementary cell to find the expression of the exact coordinates of the outgoing trajectory.  The scattering map of Eq.~\ref{eq:complete} restricted to the domain $\bigcup_{i=0}^{11} \beta_i$, is given by
\begin{equation}\label{eq:Sg0}
    \fl  \mathcal{S}_{1/2,\pi/2}(h, \theta) = \left\{
    \begin{array}{ll}
      (-h \cot \theta, \frac{\pi}{2} - \theta)					       & (h, \theta) \in \beta_0 \\ 
      (h + \frac{1}{2}\left[ \alpha(1 + \tan \theta) + 3 \tan\theta + 1 \right], \theta) & (h, \theta) \in \beta_1 \cup \beta_4 \\ 
      ((h + \frac{1}{2})\cot \theta + \frac{5}{2}, -\theta - \frac{\pi}{2})	       & (h, \theta) \in \beta_2 \\
      (h + \frac{1}{2} (1 + 5 \tan \theta), \theta)			       	       & (h, \theta) \in \beta_3 \\
      (h \cot \theta + 2d + 1, -\theta - \frac{\pi}{2})			       	       & (h, \theta) \in \beta_5 \\
      (h + (1 + 2d) \tan \theta, \theta)					       	       & (h, \theta) \in \beta_6 \\
      (h + \frac{3}{2}\tan\theta - \frac{1}{2}, \theta)			       	       & (h, \theta) \in \beta_7 \\
      (1 - (h + (1 + 2d) \tan\theta), \theta + \pi)			       	       & (h, \theta) \in \beta_8 \cup \beta_9 \\
      ((h - 1) \cot \theta + 1 + 2d, \frac{3\pi}{2} -\theta)		       	       & (h, \theta) \in \beta_{10} \\
      (-h + (1 + d)(1 - \tan \theta), \theta + \pi)           	       	       & (h, \theta) \in \beta_{11}
    \end{array} \right.
\end{equation}
We have seen that in some regions, the existence of bouncing trajectories has to be taken with special care to identify the right coordinates of the outgoing trajectory. Also, even in certain cases, different regions corresponding to different itineraries may have the same coordinates for the outgoing trajectory, as is the case for $\beta_8$ and $\beta_9$ seen in Eq.~\ref{eq:Sg0}.

To end this section, in Table.~\ref{tab:iting0} we list the symbolic itineraries corresponding to incoming trajectories having a first collision with the boundary $\G_0$.

\begin{table}[!ht]
  \centering
  \begin{tabular}{c|l}
    \hline
    \textbf{Domain} & \textbf{Symbolic itinerary} \\
    \hline 
    $\beta_0$			   & $\mathcal{L} \G_0 \mathcal{L}$ \\
    $\beta_1 \cup \beta_4$ & $\{\mathcal{L} \G_0 \G_3 \mathcal{R}, \mathcal{L} \G_0 \G_3 \G_0 \G_3 \mathcal{R}, ... \}$ \\
    $\beta_2$			   & $\mathcal{L} \G_0 \G_3 \G_0 \G_3 \G_2 \mathcal{R}$ \\
    $\beta_3$ 			   & $\mathcal{L} \G_0 \G_3 \G_0 \G_3 \G_2 \G_1 \mathcal{R} $ \\
    $\beta_5$ 			   & $\mathcal{L} \G_0 \G_3 \G_2 \mathcal{R}$ \\
    $\beta_6$ 			   & $\mathcal{L} \G_0 \G_3 \G_2 \G_1 \mathcal{R}$ \\
    $\beta_7$ 			   & $\mathcal{L} \G_0 \G_3 \G_2 \G_1 \G_2 \G_1 \mathcal{R}$ \\
    $\beta_8$			   & $\mathcal{L} \G_0 \G_3 \G_2 \G_0 \mathcal{L}$ \\
    $\beta_9$			   & $\mathcal{L} \G_0 \G_2 \G_3 \G_0 \mathcal{L}$ \\
    $\beta_{10}$		   & $\mathcal{L} \G_0 \G_2 \G_3 \mathcal{L}$ \\
    $\beta_{11}$ 		   & $\mathcal{L} \G_0 \G_2 \mathcal{L}$ \\
  \end{tabular}
  \caption{Table of the symbolic itineraries for trajectories in the partition regions $\{\beta_j\}$.}
  \label{tab:iting0}
\end{table}

\subsection{A first collision with the boundary $\Gamma_2$}
\label{sec:g2}

\begin{figure}[!t]
    \centering
    \includegraphics[width=0.7\textwidth]{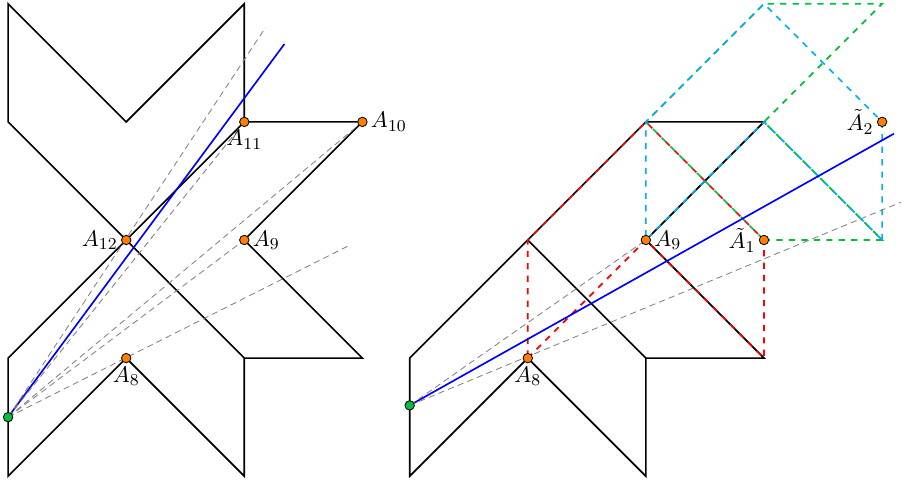}
    \caption{Translation surface of the elementary cell for incoming rays that first hit the edge $\G_2$. Left panel: Unfoldings used to define the splitting of directions, $\rho_m$ for $m = 9,10,11,12$.  Right panel: Unfolding along a direction $\theta \in \rho_8$. The vertices $\widetilde{A}_1, \widetilde{A}_2$ determine the nested singular directions within $\rho_8$.}
    \label{fig:g2unfold}
\end{figure}

In this section we obtain the scattering map $\mathcal{S}_{1/2,\pi/2}(h,\theta)$ for trajectories restricted to have a first collision with the boundary $\G_2$.  Proceeding as in Sec.~\ref{sec:g0}, we first determine a minimal covering of the restricted domain by identifying the singular directions $\xi_m$, from the unfolding of the elementary cell into a translation surface. This is schematically shown in left panel of Fig.~\ref{fig:g2unfold}, where five such singular directions are identified by connecting the initial coordinates of the incoming ray (green circle) with the accessible vertices $A_8,\ldots A_{12}$ (dashed gray lines).

For these trajectories, the minimal partition is comprised of four regions, that we denote as  $\beta_{i+3} = \mathcal{D} \times \rho_{i}$, for $i \in \{9,10,11,12\}$. The regions $\beta_{13}$, $\beta_{14}$ and $\beta_{15}$ do not contain any nested singular trajectories and thus, the calculation of the scattering map in these regions is straightforward.

In the region $\beta_{12}$, we find a family of bouncing trajectories characterized by the words $\{ \lb\G_2 \G_1\rb, \lb\G_2 \G_1 \G_2 \G_1\rb, \ldots \}$. From the translation surface shown in the right panel of Fig.~\ref{fig:g2unfold}, we identify the singular directions $\widetilde{\xi}_n(h)$ defined by the vertices $\widetilde{A}_n = (1 + nd, 1 + (n-1)d)$, for $n=\{0,1,\ldots\}$. Note that $\widetilde{\xi}_0$ is an inaccessible singular direction that is included here for convenience.

The regions defined by $\widetilde{\xi}_n$ are determined by intersecting $\widetilde{\xi}_n(h)$ with both $\xi_{8}(h)$ and $\xi_9(h)$. We obtain an infinite sequence of nested domains for the height coordinate of the incoming ray:  $\widetilde{D}_1 = \left(\frac{1}{4}, d \right)$ and $\widetilde{D}_n = \left( \frac{1}{2(n+1)}, \frac{1}{n}\right)$ for $n \ge 2$. Since $\widetilde{D}_n$ is not bounded, we redefine these regions as $\beta_{12}^{(n)} = \left(\mathcal{D} \times \widetilde{\rho}_n(h)\right) \cap \beta_{12}$ where $\widetilde{\rho}_n(h) = \{ \theta : \ \widetilde{\xi}_n(h) < \theta < \widetilde{\xi}_{n+1}(h)\}$.

Similarly to the family of bouncing trajectories found in Sec.~\ref{sec:g0}, the area of $\beta_{12}^{(n)}$ shrinks as $(h, \theta) \to (0, \pi/4)$. By applying the same reasoning as in Sec.~\ref{sec:g0}, we derive the inequality $y(a_n) \le b_{n+1}$ where $a_n = 1 +n d$ and $b_n = (1+n)d$ are the vertical and horizontal levels of the step function that envelope the relevant opening segments of the translation surface. Solving for the index $n$ we find
\begin{equation} \label{eq:iotag2}
  n \le \frac{d - h - \tan \theta}{d(\tan \theta - 1)} \ , \quad
  \iota(h,\theta) = \left\lceil \frac{d - h - \tan \theta}{d(\tan \theta - 1)} \right\rceil \ .
\end{equation}
In general, $\iota(h,\theta) = n$ for $(h,\theta) \in \beta_{12}^{(n)}$. The right panel of Fig.~\ref{fig:g2unfold} corresponds to $\iota = 2$.

Finally, solving for the intersection points $P,P'$ as before, the scattering map restricted to rays first hitting the boundary $\G_2$ is
\begin{equation} \label{eq:sg2}
    \fl    \mathcal{S}_{1/2,\pi/2}(h, \theta) =  \left\{
    \begin{array}{ll}
	(h + (1 + d \iota) \tan \theta - d \iota, \theta)         & (h,\theta) \in \beta_{12} \\
	    (-h + (d + 1)(1 -  \tan \theta), \theta + \pi)	        & (h,\theta) \in \beta_{13} \\
	    (-h \cot \theta + (d+1)(\cot \theta - 1), \frac{3\pi}{2} - \theta)& (h,\theta) \in \beta_{14} \\
	    (-h - \tan \theta + 2d + 1, \theta +\pi)		        & (h,\theta) \in \beta_{15} \\
    \end{array} \right. \ .
\end{equation}
A precise definition of the partition regions $\beta_m$ is given in App.~\ref{app:A}.
The symbolic itineraries characterizing each region are listed in Table~\ref{tab:iting2}.
\begin{table}[h]
  \centering
  \begin{tabular}{c|l}
    \hline
    \textbf{Domain} & \textbf{Itinerary} \\
    \hline 
    $\beta_{12}$ & $\{\mathcal{L} \G_2 \G_3 \mathcal{R}, \mathcal{L} \G_2 \G_3 \G_2 \G_3 \mathcal{R}, \ldots \}$ \\
    $\beta_{13}$ & $\mathcal{L} \G_2 \G_0 \mathcal{L}$ \\
    $\beta_{14}$ & $\mathcal{L} \G_2 \mathcal{L}$ \\
    $\beta_{15}$ & $\mathcal{L} \G_2 \G_3 \mathcal{L}$ \\
  \end{tabular}
  \caption{Table of itineraries for trajectories associated with regions $\beta_j$ where $j = 12,13,...,15$.}
  \label{tab:iting2}
\end{table}

\subsection{A first collision with the boundary $\G_3$}
\label{sec:g3}

\begin{figure}[!t]
    \centering
    \includegraphics[width=0.7\textwidth]{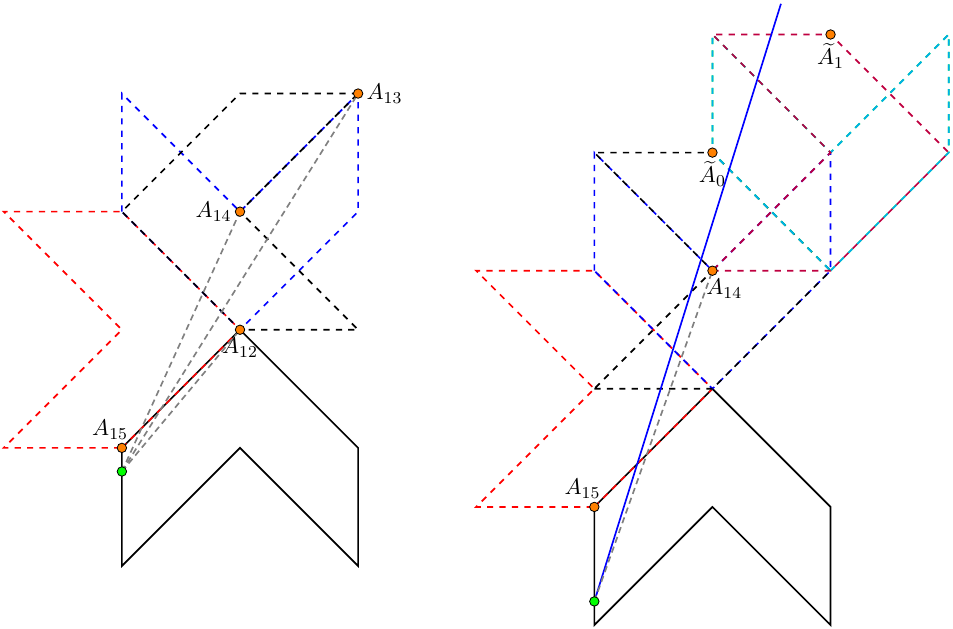}
    \caption{Translation surface of the elementary cell for incoming rays that hit the edge $\G_3$ first.  Left panel: The singular directions given by the accessible vertices $A_m$ for $m = 12,13,...,15$ are indicated by the gray dashed lines. Right panel: Singular directions associated to bouncing trajectories in the interval $\beta_{18}$, as indicated by the vertices $\widetilde{A}_n$}
    \label{fig:g3unfold}
\end{figure}

To complete the derivation of the scattering map $\S_{1/2,\pi/2}$ of the cell with finite horizon, we need to analyze the trajectories that have initial conditions in $\B^+$ and have a first collision with the boundary $\G_3$. The translation surface obtained from unfolding the elementary cell is shown in the left panel of Fig.~\ref{fig:g3unfold}. In this case we obtain three singular directions $\xi_m(h)$ with $m = 13, 14, 15$ defined by joining the initial coordinates of the incoming ray with the position of the corresponding vertices $A_m$. Notice that the vertex $A_{12}$ has already been considered in Sec.~\ref{sec:g2}.

As before, we define a corresponding minimal partition $\beta_{m+3} = \mathcal{D} \times \rho_{m}$, continuing the enumeration of previous sections. We find that the regions $\beta_{16}$ and $\beta_{17}$ do not contain any further singular trajectories and for these partition regions the derivation of the scattering map is straightforward. We focus instead on the region $\beta_{18}$ which contains a nested singular direction and the corresponding bouncing trajectories whose symbolic itineraries are $\{ \lb \G_3 \G_2 \G_1 \rb, \lb \G_3 \G_2 \G_1 \G_2 \G_1 \rb\}$ (see the right panel of Fig.~\ref{fig:g3unfold}). Different from the previous cases, this family of bouncing trajectories has just two members. We have found that the number of members and thus the number of nested singular directions, increases in the expected way as the channel becomes more narrow, $d\rightarrow0$.

Proceeding as before, the index function is 
\begin{equation} \label{eq:iotag3}
    \iota(h, \theta) = \left\lceil \frac{h - 1 - d}{d(1 - \tan \theta)} \right\rceil \in \{1,2\}
\end{equation}
and is obtained by solving $b_n \le y(a_n)$ for $n$, where $(a_n, b_n) = (n d, 1 + (1 + n)d)$. Evidently, $\iota$ takes on just two values, meaning that $\beta_{18}$ is divided into two regions, defined in App.~\ref{app:A}.

The obtained scattering map $\mathcal{S}_{1/2,\pi/2}$ is 
\begin{equation} \label{eq:sg3}
    \fl   \mathcal{S}_{1/2,\pi/2}(h, \theta) = \left\{
    \begin{array}{ll}
	(1 + 2d - (h + \tan \theta), \theta + \pi)		       & (h,\theta) \in \beta_{16} \\
	(-\cot \theta (1 + 2d - h) + 1, \frac{3 \pi}{2} - \theta)  & (h,\theta) \in \beta_{17} \\
	(\cot\theta  (-h + \frac{1}{2}(\iota + \tan \theta ( 1 - \iota) + 3)), \frac{\pi}{2} - \theta) & (h,\theta) \in \beta_{18}
    \end{array} \right. \ ,
\end{equation}
and the corresponding symbolic itineraries are summarized in the following Table.~\ref{tab:iting3}.

\begin{table}[h]
  \centering
  \begin{tabular}{c|l}
    \hline
    \textbf{Domain} & \textbf{Itinerary} \\
    \hline 
    $\beta_{16}$ & $\mathcal{L} \G_3 \G_2 \mathcal{L}$ \\
    $\beta_{17}$ & $\mathcal{L} \G_3 \G_2 \G_0 \mathcal{L}$ \\
    $\beta_{18}$ & $\{\mathcal{L} \G_3 \G_2 \G_1 \mathcal{R}, \mathcal{L} \G_3 \G_2 \G_1 \G_2 \G_1 \mathcal{R} \}$ \\
  \end{tabular}
  \caption{Table of itineraries for trajectories associated with regions $\beta_j$ for $j = 16,17,18$.}
  \label{tab:iting3}
\end{table}

\subsection{The scattering map $\S_{1/2,\pi/2}$}

\begin{figure}[!t]
    \centering
    \includegraphics[width=0.8\textwidth]{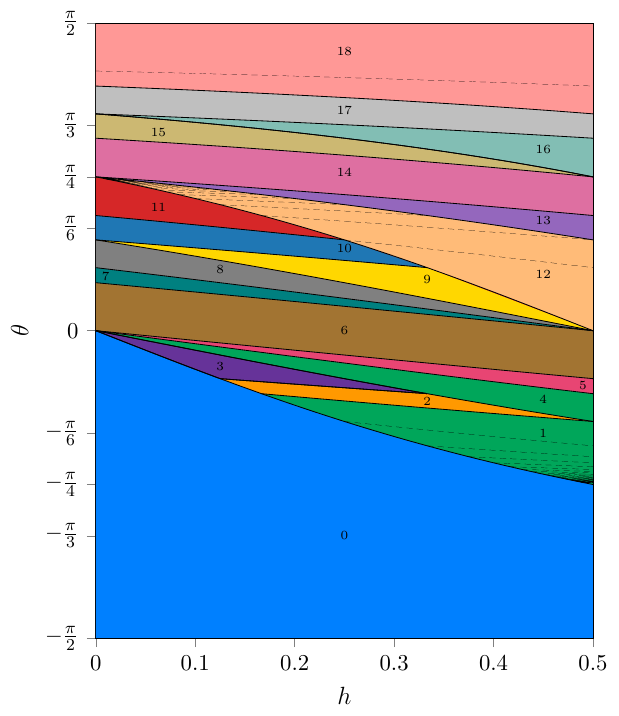}
    \caption{Partition of the coordinate space $(h,\theta)$ for an elementary cell with finite horizon, $d = 1/2$. 
      Each colored area is assigned an index $j = 0,1,...,18$ corresponding to each $\beta_j$. Solid black lines resemble the singular directions $\xi(h)$ and $\widehat{\xi}(h)$ where in the latter case $h$ is not defined on the full domain $\mathcal{D}$. Dashed lines correspond to the singular directions $\widetilde{\xi}(h)$ of the bouncing trajectories. The bouncing trajectories accumulate at points $(0,\pi/4$ and $(1/2,-\pi/4)$, as their length diverges.  }
    \label{fig:chisFH}
  \end{figure}

We have derived an exact expression for the scattering map $\S_{1/2,\pi/2}(h,\theta)$, translating any incoming trajectory, entering the elementary cell with $d=1/2$ and $\alpha=\pi/2$ through the opening $\lb$, into a unique outgoing trajectory (see Fig.~\ref{fig:schem}). We found that $\S_{1/2,\pi/2}(h,\theta)$ is defined over a partition of the coordinate space in Birkhoff variables that we derived exactly with the help of the translation surface of the unfolded elementary cell. The partition $\{\beta_m\}$ is listed in App.~\ref{app:A} while a graphical representation can be seen in Fig.~\ref{fig:chisFH}. 

\begin{figure}[!ht]
    \centering
    \includegraphics[width=0.7\textwidth]{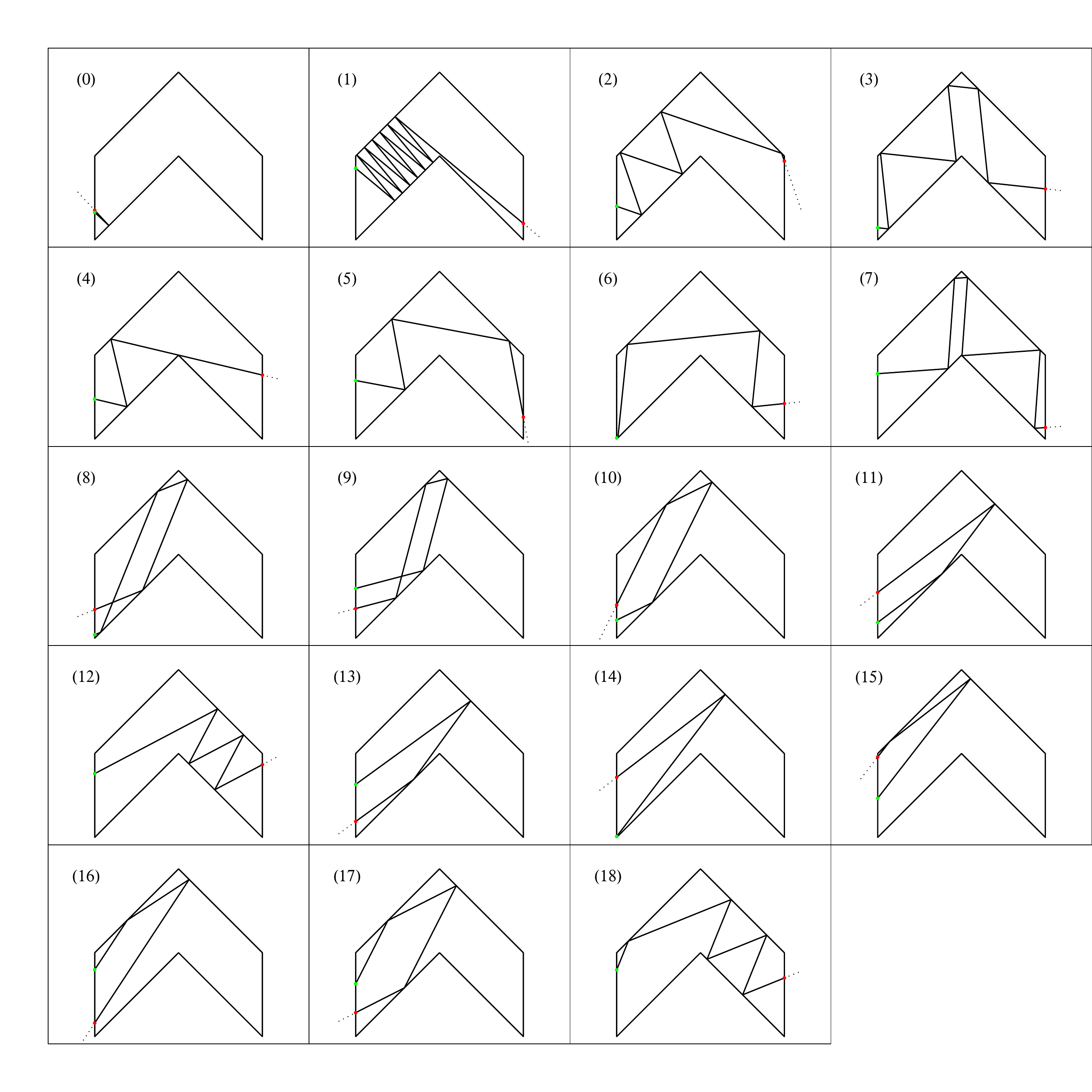}
    \caption{Geometry of the families of incoming trajectories in each of the partition elements $\beta_j$ where $j = 0,1,...,18$, for elementary cell with finite horizon, $d = 1/2$. Note that only one member of each family of bouncing trajectories is shown at the respective panels 1,12 and 18. Green and red dots respectively denote the beginning and ending of each trajectory segment.}
    \label{fig:visualsFH}
\end{figure}

Fig.~\ref{fig:chisFH} gives an indication of the intricate permutation of intervals of parameters that the billiard flow produces. The dashed curves indicate the families of bouncing trajectories. Moreover, Fig.~\ref{fig:visualsFH} shows the different type of trajectories, characterizing each region of the partition.

In the following sections \ref{sec:dx} and \ref{sec:fronts}, we will use the scattering map to efficiently solve the billiard dynamics.

\section{The scattering map for a cell with infinite horizon}
\label{sec:infinite}

In this section, we consider a polygonal billiard channel with infinite horizon, comprised by connecting side by side the elementary polygonal cell of Fig.~\ref{fig:schem} with the same opening angle $\alpha=\pi/2$ as before, but with $d=1$. With this geometry, all incoming horizontal rays entering the elementary cell through the left opening $\lb$, with coordinate $1/2<h<1$, will exit the cell through $\rb$ without colliding with the cell's boundary.  Here we carry the same analysis as in the previous section, to derive an interval exchange transformation, where the intervals in the parameter space are classified by the singular trajectories of the cell.

\begin{figure}[!t]
    \centering
    \includegraphics[width=0.7\textwidth]{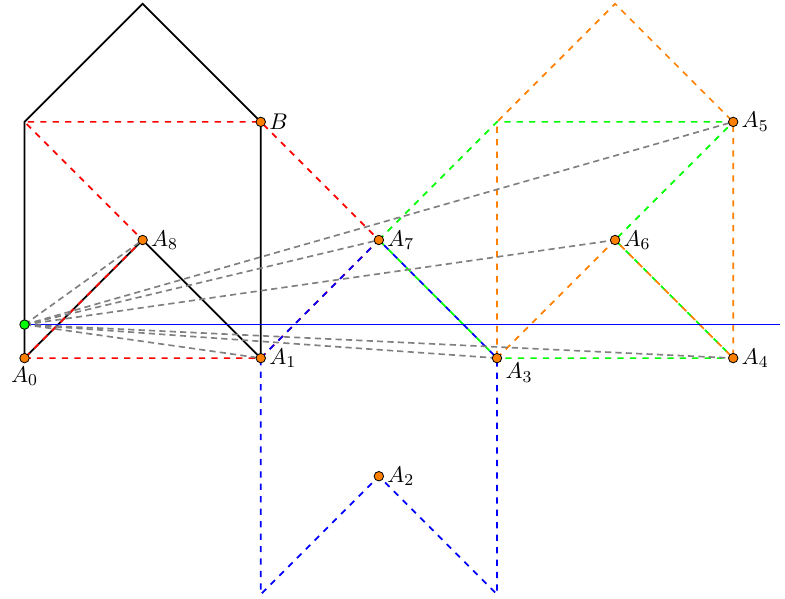}
    \caption{Translation surface of the elementary cell with infinite horizon, $d=1$, corresponding to incoming rays that hit the edge $\G_0$ first. A possible initial condition is marked by the green circle. The accessible vertices $A_m$ and $B$, appearing from the unfolding of the polygonal cell are indicated by orange circles. Note that the vertex $B$ is accessible only for cells with infinite horizon. The singular directions are indicated by the dashed gray lines.}
    \label{fig:g0unfoldIH}
\end{figure}

We start by obtaining the translation surface by unfolding the elementary cell for horizontal incoming rays by means of the Zemlyakov-Katok construction. The translation surface is shown in Fig.~\ref{fig:g0unfoldIH}. Clearly, the non trivial horizontal rays are those with $h<1/2$. Moreover, we notice that comparing the translation surface with that of the cell with finite horizon, there is one additional vertex with coordinates with coordinates $(2\delta x, d)$, that is now accessible to the flow of incoming rays. This is marked as $B$, in Fig.~\ref{fig:g0unfoldIH}.

From Fig.~\ref{fig:g0unfoldIH} we see that the order of the singular directions $\xi_m$ undergo a permutation yielding $\xi_0 \le \xi_1 \le \xi_3 \le \xi_4 \le \xi_6 \le \xi_7 \le \xi_5 \le \xi_8$, when $h < 1/2$.  Furthermore, we find that the regions $\beta_{1,2,3}$ vanish because no ray can pass between the points $A_1$ and $A_2$.  A similar situation is found for $\beta_7$ where the direction $\widehat{\xi}_3$ (top right panel of Fig.~\ref{fig:g0unfold}) no longer corresponds to an accessible singular point. Therefore, apart from $\beta_0$, all regions of the minimal partition, corresponding to incoming rays from the left that have a first collision with the boundary  $\G_0$ are in the domain $A = \mathcal{D} \times \{ \theta : \ \xi_1 < \theta < \xi_8 \}$.

For incoming rays hitting the boundaries $\G_2$ or $\G_3$ first, we realize that Equations \ref{eq:sg2} and \ref{eq:sg3} can be extended to the case $d>1/2$, in particular the present case $d=1$, without further modification.  Moreover, because of the infinite horizon, all the sub regions in $A$ are truncated by the additional region $\beta_{19} = \mathcal{D} \times \{ \theta : \ \xi_8 < \theta < \eta \}$, with $\eta$ the singular direction corresponding to the newly accessible vertex $B$. This region corresponds to the trivial rays with itinerary $\mathcal{L}\mathcal{R}$.

\begin{figure}[!t]
    \centering
    \includegraphics[width=0.8\textwidth]{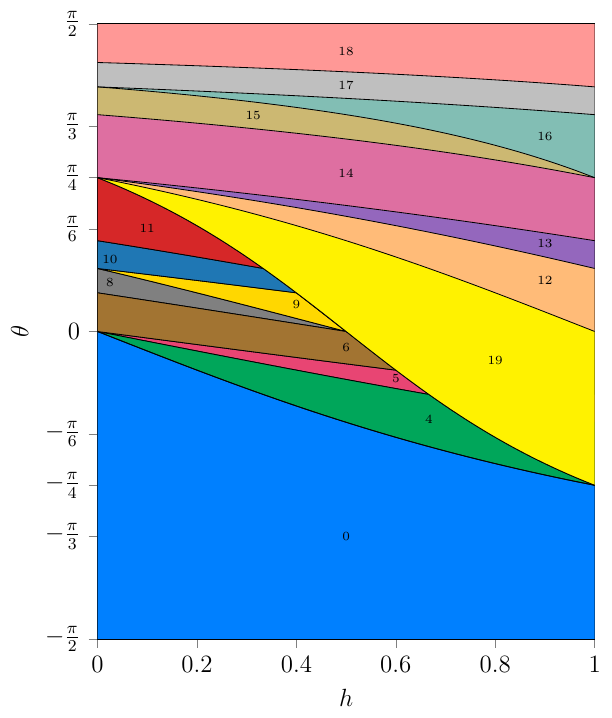}
    \caption{Partition of the coordinate space $(h,\theta)$ for elementary cell with finite horizon, $d = 1/2$.  The colour scheme is kept consistent with that of Fig.~\ref{fig:chisFH}, meaning that the symbolic trajectories found for both $d = 1$ and $d = 1/2$ are labeled equally.}
    \label{fig:chisIH}
  \end{figure}

Taking into account the permutation of the singular directions $\{\xi_m\}$, we can define sub regions of $\B^+$ as $\Z^+_{J} = \left(\mathcal{D}_{J(k)} \times \mathcal{A}_{J(k)}^+\right)\cap A$, where the vector $J$, with values $J(k) = (1,3,4,6,7,5,8)$ for $k = 0,1,\ldots,6$, takes care of the permutation. The definitions of these regions are explicitly given in App.~\ref{app:A}. Fig.~\ref{fig:chisIH} shows the partition graphically. 

We remark  that each $\Z^+_{J(k)}$ corresponds exactly to a region $\beta_j$ of the previous Sec.~\ref{sec:finite}, adjusted to  $d = 1$ and thus the expression of the scattering map $\S_{1,\pi/2}$ may be carried over from Sec.~\ref{sec:finite}. This correspondence is determined by the symbolic itineraries associated with the regions $\Z^+_{J(k)}$, as shown in the following Table.~\ref{tab:itinIHg0}, where the new emerging region due to the trajectories crossing the cell without collisions $\beta_{19}$ has been included. The rest of the itineraries can be extracted from  Tables.~\ref{tab:iting0},\ref{tab:iting2},\ref{tab:iting3}.
\begin{table}[h]
  \centering
  \begin{tabular}{c|c|l}
    \hline
    \textbf{Domain} ($d = 1/2$) & \textbf{Domain} ($d = 1$) & \textbf{Itinerary} \\
    \hline 
    $\beta_4$	  & $\Z^+_{J(0)}$ & $\lb \G_0 \G_3 \rb$ \\
    $\beta_5$ 	  & $\Z^+_{J(1)}$ & $\lb \G_0 \G_3 \G_2 \rb$ \\
    $\beta_6$ 	  & $\Z^+_{J(2)}$ & $\lb \G_0 \G_3 \G_2 \G_1 \rb$ \\
    $\beta_8$	  & $\Z^+_{J(3)}$ & $\lb \G_0 \G_3 \G_2 \G_0 \lb$ \\
    $\beta_9$	  & $\Z^+_{J(4)}$ & $\lb \G_0 \G_2 \G_3 \G_0 \lb$ \\
    $\beta_{10}$  & $\Z^+_{J(5)}$ & $\lb \G_0 \G_2 \G_3 \lb$ \\
    $\beta_{11}$  & $\Z^+_{J(6)}$ & $\lb \G_0 \G_2 \lb$ \\
    $\beta_{19}$  & $\beta_{19}$ & $\lb \rb$ 
  \end{tabular}
  \caption{Correspondence between the regions $\beta_j$ and $\Z^+_{J(k)}$ as determined by the itineraries of each region. Note that the regions $\beta_{1,2,3,7}$ vanish when setting $d = 1$.}
  \label{tab:itinIHg0}
\end{table}

As is evident from the geometry of the cell, no bouncing trajectories exist for $d\ge1$. Namely, for an arbitrary direction, a trajectory will hit the parallel bottom and top boundaries of the elementary cell only once. This fact has greatly simplified the discussion in this section. Indeed, preliminary investigations (not given here) of a scattering map with $d \in (1, \infty)$ reveal translation surfaces not dissimilar from those seen thus far. For $d<1$ the same bouncing trajectories discussed in Sec.~\ref{sec:finite} will be present.

In Eq.~\ref{eq:Sih} we explicitly define $\S_{1,\pi/2}$ for the infinite horizon cell with $d=1$. 

\begin{equation}\label{eq:Sih}
\fl    S_{1,\pi/2}(h, \theta) = \left\{
	\begin{array} {ll}
	    (-h \cot \theta, \frac{\pi}{2} - \theta) & (h,\theta) \in \beta_0 \\ 
	    (h + \frac{1}{2}\left[ 3 \tan\theta + 1 \right], \theta) & (h,\theta) \in \Z^+_{J(0)} \\
	    (h \cot \theta + 2d + 1, -\theta - \frac{\pi}{2}) & (h,\theta) \in \Z^+_{J(1)} \\ 
	    (h + (1 + 2d) \tan \theta, \theta) & (h,\theta) \in \Z^+_{J(2)} \\
	    (-h - (1 + 2d)\tan\theta + 1, \theta + \pi) & (h,\theta) \in \Z^+_{J(3)} \cup \Z^+_{J(4)} \\
	    (-(1 - h) \cot \theta + 1 + 2d, \frac{3\pi}{2} -\theta)     & (h,\theta) \in \Z^+_{J(5)} \\
	    (-h + (1 + d)(1-\tan \theta), \theta + \pi)       & (h,\theta) \in \Z^+_{J(6)} \\
	    S_{\G_2}(h, \theta) & (h,\theta) \in \bigcup_{i=12}^{15} \beta_i \\
	    S_{\G_3}(h, \theta) & (h,\theta) \in \bigcup_{i=16}^{18} \beta_i \\
	    (h + \tan \theta, \theta) & (h,\theta) \in \beta_{19}
	\end{array} \right. \ .
\end{equation}

To compare the effect that the infinite horizon has on the billiard flow, the color coding and enumeration used in Fig.~\ref{fig:chisIH} is the same as in Fig.~\ref{fig:chisFH}. A salient feature of Fig.~\ref{fig:chisIH} is the dominance of the region $\beta_{19}$ over the incoming directions $-\pi/4<\theta<\pi/4$.
We do also observe the disappearance of regions $\beta_1$, $\beta_2$ and $\beta_3$ in favor of the region $\beta_0$ that corresponds to the incoming rays that are reflected at the boundary $\G_0$ back to the left opening $\lb$.

To end this section, in Fig.~\ref{fig:visualsFH} we show the different type of trajectories characterizing each region in the partition.
  
\begin{figure}[!t]
    \centering
    \includegraphics[width=0.7\textwidth]{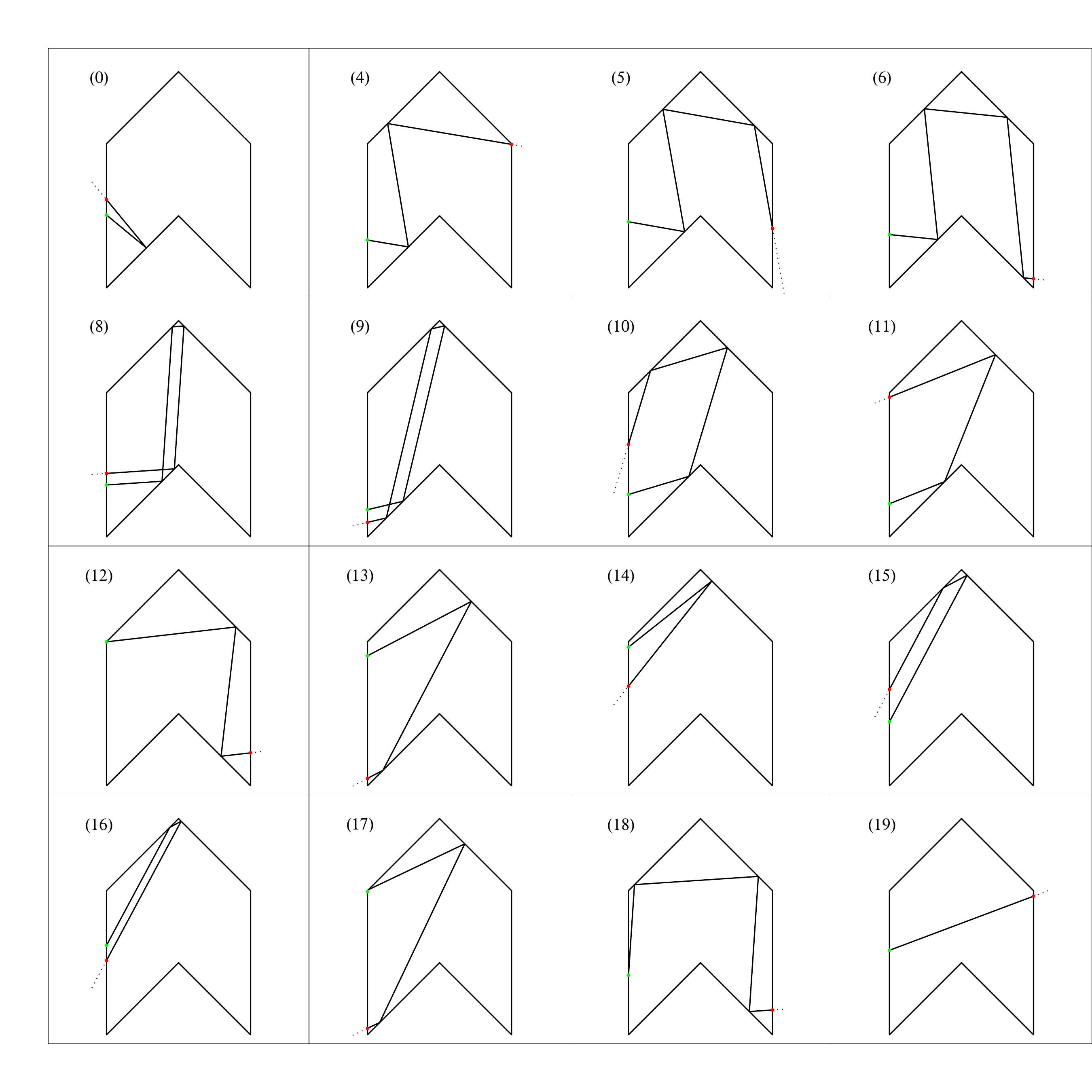}
    \caption{Geometry of the families of incoming trajectories for elementary cell with infinite horizon, $d = 1$.}
    \label{fig:visualsIH}
\end{figure}

\section{Particle transport and ballistic fronts}
\label{sec:dx}

Armed with the scattering map $\widehat{\mathcal{S}}_{d,\alpha}$, in this section we looked at the particle transport in a polygonal channel consisting of an infinite sequence of elementary cells connected through their openings. Repeated iterations of the scattering map yield the position $x$ of the particle along the channel. Consider a particle with initial condition $x_0=0$, $y_0=h$ and a unit velocity vector with direction $\theta_0 \in [0, 2\pi)$. After one iteration of the scattering map we obtain
\begin{equation} \label{eq:iter}
  (y_1,\theta_1) = \widehat{\mathcal{S}}_{d,\alpha}(y_0,\theta_0) \ , 
\end{equation}
and the position along the channel $x_1=\{x_0, \pm 2\delta x\}$ depends on the final direction of the velocity vector. If $x_1=x_0$, then it means that its path has been reflected. The other two values correspond to transmission to the right if $-\pi/2<\theta_0<\pi/2$, or to the left if $\pi/2<\theta_0<3\pi/2$.

The position of the particle after $n$ iterations of the scattering map can be written as
\begin{equation}
  \label{eq:xn}
    x_{n} = \left\{
    \begin{array} {ll}
	x_{n-1} + 1 & \mathrm{if} \ \ v_{x;n-1} \ge 0 \ \mathrm{and} \ v_{x;n} \ge 0 \\
	x_{n-1} - 1 & \mathrm{if} \ \ v_{x;n-1} < 0  \ \mathrm{and} \ v_{x;n} < 0 \\
	x_{n-1}     & \mathrm{otherwise}
    \end{array} \right. \ ,
\end{equation}
where $v_{x;n}=\cos \theta_{n}$ is the $x$-component of the particle's velocity after $n$ scatterings.  Note that $x_n$ takes on integer values, in one-to-one correspondence with the cell number, if the latter are also labeled on the integers.

\subsection{Dwell time}
\label{sec:dwell} 

The time $\tau$ spent by the particle inside the cell can also be computed analytically. Given the constant particle speed, that we set to $|\vec{v}|=1$, $\tau$ equals the length of the trajectory which can be easily obtained from the geometry of the translation surfaces. We obtain 
\begin{eqnarray} 
    \tau_{0,5} & =       -h \csc \theta			 \nonumber \\
   \tau_1 	    &= d (\iota + 3) \sec \theta \nonumber \\
   \tau_2 	    &= -(h + d) \csc \theta	 \nonumber \\
   \tau_{3,7}	    &= (3/2 + 2d) \sec \theta    \nonumber \\
   \tau_{4,11,13}  &= (d + 1) \sec \theta	 \nonumber \\
   \tau_{6,8,9}    &= (1 + 2d) \sec \theta	 \nonumber \\
   \tau_{10}	    &= (1 - h) \csc \theta	 \label{eq:tau} \\
   \tau_{12}	    &= (1 + \iota d) \sec \theta \nonumber \\
   \tau_{14} 	    &= (d + 1 - h) \csc \theta   \nonumber \\
   \tau_{15,16,19}      &=   \sec \theta	 \nonumber \\
   \tau_{17}	    &= (1 + 2d) \csc\theta       \nonumber \\
   \tau_{18} 	    &= (1 + d(1 + \iota)) \csc \theta \nonumber
 \end{eqnarray}
 where the time $\tau_n$ denotes the time that a particle, with incoming coordinates in the region $\beta_n$, spends inside the cell. In Eq.~\ref{eq:tau},
 the index function $\iota$ is given by Eq.~\ref{eq:iotag0} for $\tau_2$, Eq.~\ref{eq:iotag2} for $\tau_{12}$, and Eq.~\ref{eq:iotag3} for $\tau_{18}$.
With the exception of $\tau_{19}$ that correspond to the $\lb\rb$ trajectories which exist only with infinite horizon $d>1/2$, all dwell times are well defined for both $d = 1/2$ and $d = 1$.
Notice that different families of regions given by the indices $\{\{3,7\}, \{4,13,11\}, \{ 6,8,9 \}, \{ 15,16 \} \}$ possess times which are functionally equivalent. As discussed in Sec.~\ref{sec:g0}, this is due to the unfolded exit segments $\mathcal{L}$ and $\mathcal{R}$ having the same horizontal or vertical coordinate for different regions.

In scattering theory, the times $\tau$ correspond to the dwell time, defined as the amount of time spent by a particle inside a given coordinate range (typically a potential barrier or interaction region) \cite{smith1960}. Dwell times are an important observable used to characterize and reconstruct the scattering potential and are often experimentally observable. Diverging dwell times result from incoming trajectories approaching periodic orbits of the scattering potential. Studying the hierarchy of such divergences yield valuable information about the hierarchy of the periodic orbits of the potential. For instance, an intricate hierarchy characteristic of chaotic scattering potentials can be used as an external ``microscope'' to reconstruct the otherwise inaccessible chaotic saddle \cite{jung1999}, although in some situations, using scattering functions to distinguish chaos from integrable motion may be cumbersome \cite{jung1995}.

\begin{figure}[!t] 
    \centering
    \includegraphics[width=\textwidth]{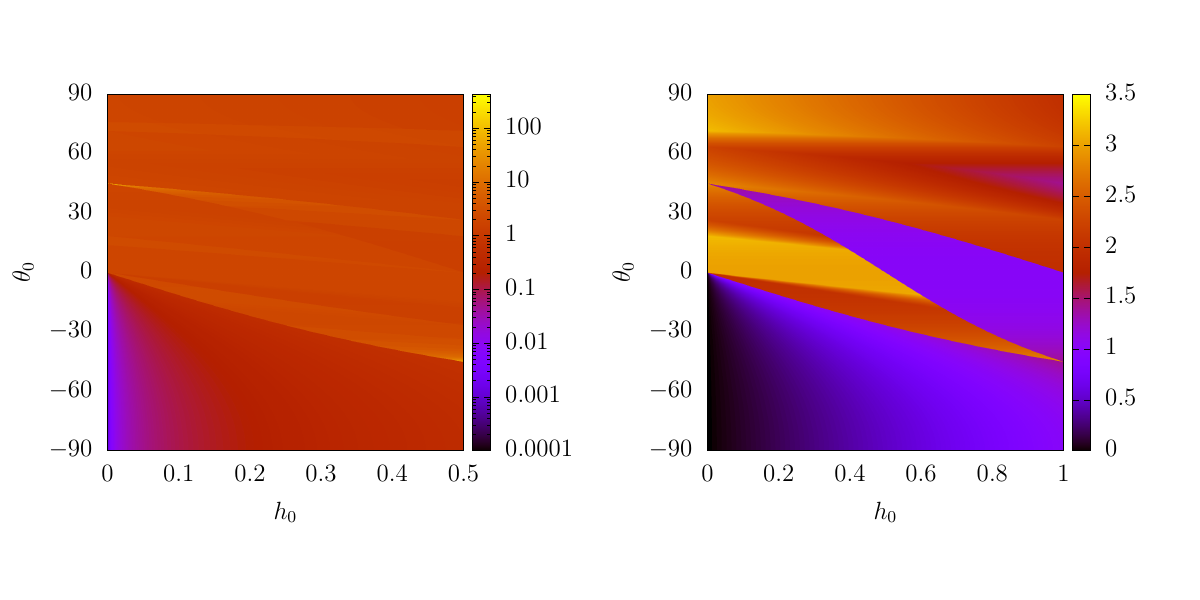}
    \vspace{-1cm}
    \caption{Heat map of the dwell time $\tau_j(h, \theta)$ of Eq.~\ref{eq:tau}, for the finite horizon cell (left) and infinite horizon cell (right).} 
    \label{fig:tauheatmap}
\end{figure}

The scattering function of dwell times $\tau(h_0,\theta_0)$, given by Eq.~\ref{eq:tau}, is the time $\tau$ spent by an incoming trajectory as a function of its initial conditions $(h_0,\theta_0)$, parametrising the impact parameter. We show in Fig.~\ref{fig:tauheatmap}, $\tau(h_0,\theta_0)$ for the cell with finite horizon (left panel for $d=1/2$), and the cell with infinite horizon (right panel for $d=1$). We notice immediately the similarity of the structure of the scattering function with the corresponding partitions of the coordinate space, Figs.~\ref{fig:chisFH} and \ref{fig:chisIH}. For the cell with finite horizon, we see the divergence of the dwell time due to the families of bouncing trajectories that approach the periodic orbits located at $(1/2,-\pi/4)$, and $(0,\pi/4)$. For the cell with infinite horizon, we see the appearance of the region $\beta_{19}$, characterized by a constant dwell time $\tau=1$. Note the difference in magnitude between the dwell times of the two cells.


\subsection{Ballistic fronts}
\label{sec:fronts} 

In Ref.~\cite{orchard2021}, we observed ballistic fronts appearing in the tails of $P(x,t)$ with a location in scaled coordinates that we denoted as $v_{\mathrm{bal}} = \Delta x / t$. These ballistic signatures are the result of families of trajectories that travel with constant speed $\nu$ along the $x$-axis of the channel, $x(t) = \nu t$. While these ballistic modes are important in a series of experiments, \emph{e.g.}, to study transmission in finite channels, in infinite channels the ballistic modes decay rapidly and therefore, do not contribute to the asymptotic statistics of the particle displacement.

Ballistic modes appear naturally in billiards with infinite horizon. For the polygonal cell of Sec.~\ref{sec:infinite}, trajectories $\lb\rb$ of region $\beta_{19}$ are ballistic. Surprisingly, in Ref.~\cite{orchard2021} we found not only that ballistic modes persist in polygonal billiards with finite horizon, but also that fronts with ballistic displacement dominate the tails of $P(x,t)$. Some indications of the existence of ballistic fronts were already observed in \emph{e.g.}, Ref.~\cite{jepps2006}. Also in Ref.~\cite{vollmer2021} it was shown that for polygonal channels with parallel boundaries, while the mean square displacement of the particle determines the statistics of the bulk of the distribution $P(x,t)$, higher moments exhibit a ballistic scaling, namely $\langle x(t)^n \rangle \sim t^{n}$, a situation known as strong anomalous diffusion \cite{castiglione1999}.

In this section we study the relation between the speed of ballistic fronts $v_{\mathrm{bal}}$ and families of propagating trajectories; trajectories that persistently move in the same direction.  We define the speed of a propagating trajectory as
\begin{equation}\label{eq:nu}
    \nu = \frac{2\delta x \cdot k }{\sum_{i\in \mathcal{I}} \tau_i} \ , 
\end{equation}
where $k>0$ is the number of cells traversed during a propagation cycle. The sum in Eq.~\ref{eq:nu} is taken over a set $\mathcal{I}$ having $k$ elements, each pointing to a dwell time of Eq.~\ref{eq:tau}. In this way, Eq.~\ref{eq:nu} accounts for propagating trajectories that pass through one or more regions in the partition $\{ \mathcal{B}_j \}$. Note that the speed of propagation does not coincide with the $x-$component of the particle's velocity, but is simply the inverse of the trajectory length taken over the number of cells traversed.

For instance, consider the elementary cell with infinite horizon of Sec.~\ref{sec:infinite}, and horizontal trajectories with incoming coordinates $(h,0)$. Trajectories with the coordinate $1/2<h<1$ (in region $\beta_{19}$), traverse the cell with speed $\nu=1$, while for coordinate $h<1/2$ (in region $\beta_{6}$), the speed is $\nu=2/3$. This shows that different families of propagating trajectories in general have different speeds $\nu$. In both cases, the propagating trajectories are periodic in $k = 1$ cell and do not leave their respective regions. 

To generalize the observation above, consider incoming trajectories with initial conditions in $(h, \theta) \in \mathcal{D} \times \{0,\pi\}$,  and an elementary cell with arbitrary width $d$. These orbits belong to region $\beta_6$. The speed of propagation becomes
\begin{equation}\label{eq:nu2}
    \nu = \frac{2\delta x}{1 + 2d n(h)} \ ,
\end{equation}
where $n(h)$ is an integer function equal to one half the number of times the trajectory hits the upper boundary of the elementary cell. The value of $n(h)$ can be written as
\begin{equation}\label{eq:n0}
    n(h) = \left\lceil \frac{\delta y - h}{d} \right\rceil = \left\{
    \begin{array} {ll}
	    N(d)+1  \ ,&\mathrm{if} \ h \in (0, \delta y - N d  ) \\
	    N(d)    \ ,&\mathrm{if} \ h \in (\delta y - N d, d) 
    \end{array} \right. \ ,
\end{equation}
where $\delta y =1/2$, and $N(d) = \left\lfloor \frac{\delta y}{d}  \right\rfloor$, where $\lfloor\cdot\rfloor$ is the greatest integer function.

Noticing that for any $h\in\D$, $n(h)$ takes only two values, then assuming that the initial coordinate $h$ of the incoming trajectories is uniformly distributed, the average speed of propagation becomes
\begin{eqnarray} \label{eq:vbalexact}
    \overline{\nu}(r) &= \frac{1}{d} \left [\frac{\delta y - N d}{1 + 2d(N+1)}  + \frac{d - \delta y + Nd}{1 + 2dN} \right] \nonumber \\
		      &= \frac{r(2N + 1)}{(rN + 1)(r(N + 1) + 1)}  \ ,
\end{eqnarray}
where  $r = d/\delta y$, so that $r>1$ corresponds to infinite horizon.

\begin{figure}[!t] 
    \centering
    \includegraphics[width=0.7\textwidth]{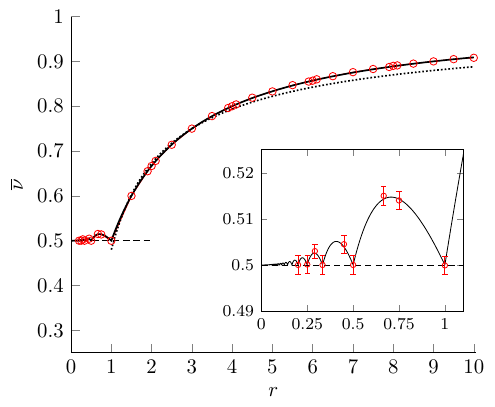}
    \caption{Average propagation speed $\overline{\nu}(r)$ as a function of the ratio $r=d / \delta y$ (solid curve), as given by Eq.~\ref{eq:vbalexact}. The red circles corresponds to the numerically obtained speed of the sub-leading ballistic front $v_{\mathrm{bal}}$.   The dotted line corresponds to the fitting curve $1 - 0.52 r^{-2/3}$ used previously in Ref.~\cite{orchard2021}. The inset shows a blown up of the region $r < 1$. } 
    \label{fig:vbal}
\end{figure}

The analytical expression Eq.~\ref{eq:vbalexact}, is a direct result of the scattering map through the dwell time of Eq.~\ref{eq:tau}. In Fig.~\ref{fig:vbal} we compare Eq.~\ref{eq:vbalexact} as a function of $r$, with the speed of the sub-leading ballistic front (red circles). In the previous Ref.~\cite{orchard2021}, we studied the ballistic components appearing in the tails of the probability distribution of the particle's displacement $P(x,t)$ and discussed the dependence of propagation speed, which we called ballistic fronts, on the geometry of the polygonal cell. In Ref.~\cite{orchard2021} a ballistic front was defined as a signature in the probability distribution of the displacement satisfying a ballistic scaling $P(x,t) = t \mathcal{F}(x/t)$. Then, we defined the sub-leading ballistic front as the fastest ballistic front, not including the uncoupled front with speed $v_{\mathrm{bal}}=x/t=1$. For more details on the numerical analysis of the ballistic fronts, we refer the reader to Ref.~\cite{orchard2021}.

The agreement of the numerical data with Eq.~\ref{eq:vbalexact} is remarkable. For infinite horizon, the average propagation speed grows and converges in the large-$r$ limit to the maximal speed $\overline{\nu}=1$. For finite horizon $r\le1$,  $\overline{\nu}(r)$ shows an oscillatory behavior that arises from the oscillations of the greatest integer function $n(d)$. This suggests that the propagating trajectories form a skeleton for the statistical behavior of the tails of the distribution $P(x,t)$.

The question that arises is why the average propagation speed of Eq.~\ref{eq:vbalexact} describes the speed of the sub-leading ballistic front. In fact, the tails of $P(x,t)$ contain several ballistic signatures, each of which is characterized by a different speed.  This is not strange if we note that trajectories propagating over a large number of cells are ``periodic orbits'' of the elementary cell, and that periodic orbits in rational polygons are known to be dense \cite{boshernitzan1998}. To investigate this further, we have numerically computed the distribution $P(x,t)$ solving the evolution by means of iterating the scattering map $\widehat{\mathcal{S}}_{d,\alpha}$.

\begin{figure}[!t] 
    \centering
    \includegraphics[width=0.7\textwidth]{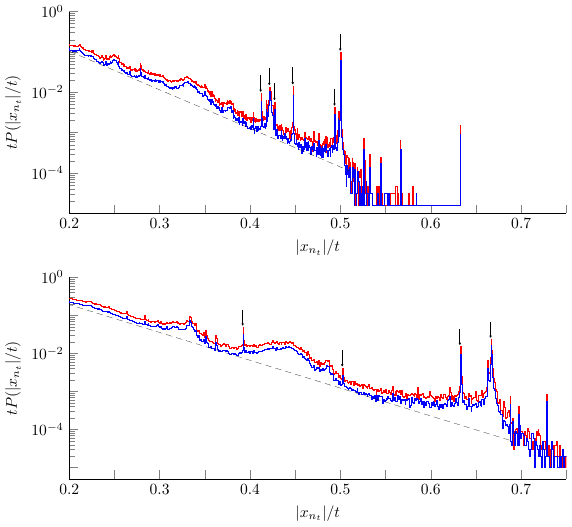}
    \caption{Distribution of the particle displacement $P(x_{n_t} ,t)$ for the finite horizon cell (top) and the infinite horizon cell (bottom). The distribution is shown for times $t = 36000$ (blue) and $t = 60000$ (red). Black arrows in both panels indicate the speed corresponding to different ballistic fronts: $v_{\mathrm{bal}} = 0.4213, 0.4473, 0.4998, 0.6321$ (top panel), and $v_{\mathrm{bal}} = 0.3328, 0.3919, 0.6321, 0.6662$ (bottom panel.} 
    \label{fig:Pxt}
\end{figure}

To study the time dependence of the particle displacement, we introduce the index $n_t = {\mathrm{arg\,min}}_n |s_n - t|$ where $s_n$ is the length of the trajectory after $n$ iterations of $\widehat{\mathcal{S}}_{d,\alpha}$.  Excluding the regions $\beta_1$ and $\beta_{12}$ to avoid bouncing trajectories, the difference $|s_{n_t} - t|$ is small, in which case $n_t$ is a good approximation of the true time. To avoid further spurious effects due to the bouncing trajectories, we have not considered realizations of the dwell time larger than the desired time $t$.

We approximate the distribution of the particle displacement as $P(x_{n_t}, t)$ for two different times $t=36000$ and $t=60000$ and $10^8$ realizations.  In Fig.~\ref{fig:Pxt} we show $P(x_{n_t}, t)$ in a ballistic scale, for the cell with finite horizon $d=1/2$ (top panel), and the cell with infinite horizon $d=1$ (bottom panel).  We see clear distinctive signatures, some of them in the form of narrow peaks (identified with arrows). 
The fact that these structures coincide at two different times over a ballistic scale means that the signatures are indeed ballistic.

Let us focus on some of these ballistic fronts. To relate them with certain family or families of trajectories we isolate the trajectories contributing to a given ballistic front, namely the trajectories with $x_{n_t}/t$ approximately equal to the propagation speed of the ballistic front. For the cell with finite horizon $d=1/2$ we analyze the ballistic fronts with speeds $v_{\mathrm{bal}} \in \{ 0.4118,0.4213,0.4270,0.4473,0.4935,0.4998 \}$. Notice the agreement between the prediction of Eq.~\ref{eq:vbalexact} $\overline{\nu}(r=1) = 1/2$, and the speed of the sub-leading ballistic front $0.4998$. For the cell with infinite horizon $d=1$ we analyze the ballistic fronts with speeds $v_{\mathrm{bal}} \in \{ 0.3919,0.5020,0.6321,0.6662\}$. Again, the prediction of the speed of the sub-leading ballistic front of Eq.~\ref{eq:vbalexact} and $\overline{\nu}(r=2) = 2/3$, is remarkable. \\

\begin{figure}[!t]
    \centering
    \includegraphics[width=1\textwidth]{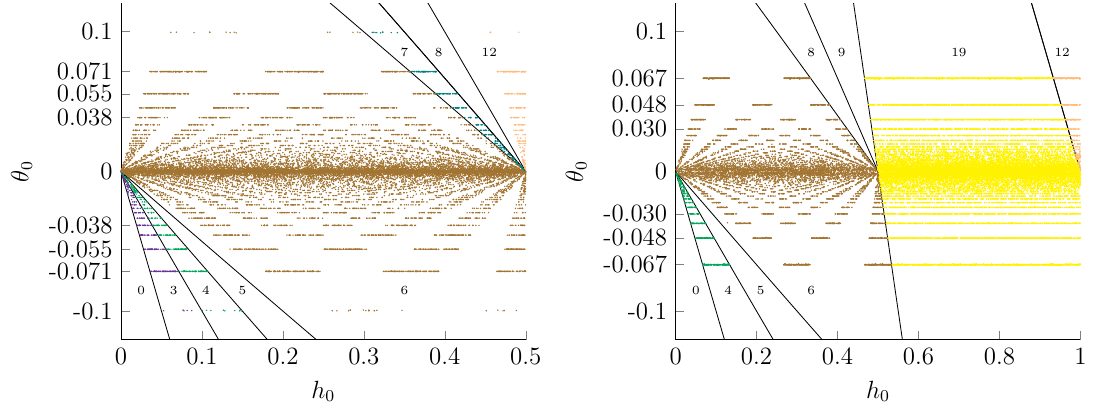}
    \caption{Initial conditions $(h_0, \theta_0)$ of trajectories contributing to the sub-leading ballistic fronts with speed $v_{\mathrm{bal}} \approx 1/2$ for the cell with finite horizon (left panel) and $v_{\mathrm{bal}} \approx 2/3$ for the cell with infinite horizon (right panel). Points and regions are colored and labeled using the same scheme as in Figs.~\ref{fig:visualsFH} and \ref{fig:visualsIH}. }
    \label{fig:vbic}
\end{figure}

To better understand the agreement between Eq.~\ref{eq:vbalexact} and the numerically obtained speeds $v_{\mathrm{bal}}$ of Fig.~\ref{fig:vbal}, we study the initial conditions of trajectories that contribute to the sub-leading ballistic front of the $d=1/2$ and $d=1$ cell. In Fig.~\ref{fig:vbic} we show a subset of the coordinate space $\B^+$ in the vicinity of $\theta_0 = 0$, with initial conditions $(h_0, \theta_0)$ corresponding to each finite (left panel) and infinite (right panel) horizon geometry. In both cases we find the sub-leading ballistic front is mostly the result of trajectories that are horizontal, near horizontal or near vertical. The latter scenario, involving initial angles $\theta_0 \approx \pm \pi/2$, has been omitted from Fig.~\ref{fig:vbic} in favour of the richer structure.

The symmetry about $\theta_0 = 0$ in Fig.~\ref{fig:vbic} is at first striking, since the polygonal cells are not horizontally symmetric. Focusing first on the cell with finite horizon, we find initial conditions in the regions $ \beta_{3}, \beta_{4}, \beta_6, \beta_7$ and $\beta_{12}$. Notably, each of these regions are transmitting, with $\theta_1 = \theta_0$. This means that the first iterate of each $(h_0, \theta_0)$ in Fig.~\ref{fig:vbic} is a horizontal shift. In fact, from the speed of the sub-leading front $v_{\mathrm{bal}} \approx 1/2$ and the definition of the dwell times Eq.~\ref{eq:tau}, it is not difficult to see that subsequent iterates must satisfy $\theta_n = \theta_0$. This is true for times $t \le 60000$ wherein the propagating trajectories with speed $\nu \approx 1/2$ have been observed. 

To give a sense of how these trajectories maintain a speed $\nu \approx 1/2$ whilst visiting multiple regions, we consider a pair of trajectories with symmetric initial conditions $(h_0, \theta_0) \approx (0.199, \pm 0.071)$, obtained from Fig.~\ref{fig:vbic}.
Evolving these points with $\widehat{\mathcal{S}}_{d,\alpha}$ allows us to study a `coordinate space itinerary' which records the regions visited by the trajectories during their lifetime. For the incoming ray with negative $\theta_0$ we define the partial itineraries $w_0 = \beta_6 \beta_3 \beta_6 \beta_6 \beta_4 \beta_6 \beta_6$ and $w_1 = \beta_6\beta_3 \beta_6$ while for its symmetric counterpart we define $u_0 = \beta_6 \beta_6 \beta_{12} \beta_6 \beta_6 \beta_7 \beta_6$ and $u_1 = \beta_6\beta_7 \beta_6$. The expressions $i)$ and $ii)$ below show the full itineraries for each case and completely specify each trajectory.\footnote{These observations are sensitive to initial conditions, requiring at least ten decimal places of accuracy. Indeed, the apparently periodic nature of $i)$ and $ii)$ does not hold at times much larger than those considered in Fig.~\ref{fig:Pxt}.}
\begin{eqnarray*}
    i) \ \theta_0 \approx -0.071: \ w_0^{912}w_1 \underbrace{w_0^{1562}w_1}_{w_2} w_2 w_2 \\
    ii) \ \theta_0 \approx +0.071: \ u_0^{650}u_1 \underbrace{u_0^{1562}u_1}_{u_2} u_2 u_2 
\end{eqnarray*}
Note that superscripts are used to denote the repetition of a given symbol, as in \emph{e.g.} $a^2 = aa$.

Let us examine the partial itineraries $w_0$ and $u_0$, which account for the majority of the propagating behavior. From Eq.~\ref{eq:nu}, we see an equivalence in the propagation speeds of $w_0$ and $u_0$ with $\nu = 7/(5\tau_6 + \tau_3 + \tau_4) = 7/(5 \tau_6 + \tau_{12} + \tau_7) \approx 0.499$. The symmetry in Fig.~\ref{fig:vbic} (left panel) therefore stems from both the equivalence of the dwell times, $\tau_3 = \tau_7$ and $\tau_4 = \tau_{12}$ with $\iota = 1$, and the propagation length $k = 7$. Notice also that $2\tau_6 = \tau_4 + \tau_3 = \tau_7 + \tau_{12}$. Thus, in general, one might expect propagation speeds of the form $\nu = k/(a\tau_6 + b\tau_{3,7} + b\tau_{4,12})$ where $k = a + 2b$ and $a,b \in \mathbb{N}$. Indeed, we reasonably suspect that similar patterns persist for smaller entrance angles, where visits to the region $\beta_6$ become more and more frequent, ultimately realizing the result of Eq.~\ref{eq:vbalexact}. 

As a final remark on the sequences $i)$ and $ii)$, we notice a transient phase involving a number of repetitions $m=912$ and $n=650$ which add to give the later observed `periodic' phase of $1562$. This may suggest an alternate symmetry involving initial conditions that lead to itineraries satisfying $m = n$.

Turning to the cell with infinite horizon $d=1$, the sub-leading ballistic front $v_{\mathrm{bal}} \approx 2/3$ is formed by trajectories with initial conditions in regions $\beta_4, \beta_6, \beta_{12}$ and $\beta_{19}$.
For the incoming trajectories in $\beta_4$ and $\beta_6$ (for  $h_0\le1/2$), we observe a similar hierarchy to the one obtained from the cell with finite horizon. For $h_0>1/2$ the hierarchy of trajectories contributing to the sub-leading ballistic front changes into discrete values of $\theta_0$. This is precisely what is expected from the incoming trajectories in region $\beta_{19}$. The discretization in the initial directions is a kind of selection rule consistent with the propagation speed of $v_{\mathrm{bal}} \approx 2/3$. Further, as with the finite horizon case, the symmetry seen in Fig.~\ref{fig:vbic} (right panel) can be understood through the equivalence in dwell times $\tau_{4} = \tau_{12}$, again with $\iota = 1$ and in addition, the relation $\tau_6 = \tau_4 + \tau_{19} = \tau_{12} + \tau_{19}$. Differently from the finite horizon case, we see that trajectories belonging to the front $v_{\mathrm{bal}} \approx 2/3$ with larger entrance angles spend the majority of their time in $\beta_{19}$ as opposed to $\beta_6$, as is evident from Fig.~\ref{fig:vbic} (right panel). Propagation in these cases is the result of long flights with intermittent visits to $\beta_4, \beta_6$ and $\beta_{12}$.  

Finally, note that, in contrast with the cell with finite horizon, horizontal rays $\theta_0 = 0$ do not contribute to the sub-leading ballistic front. Instead, these trajectories contribute to other ballistic fronts: to the front with propagation speed $\nu=1/3$ if $h_0\le1/2$, and to the front with propagation speed $\nu=1$ if $h_0>1/2$. This can be seen as the void of initial conditions along the line $\theta_0=0$. 

From the above findings we see that the average propagation speed Eq.~\ref{eq:vbalexact} correctly models the speed of the sub-leading ballistic front, even when incoming rays are not exactly horizontal, $\theta_0 \ne 0$. This is because near horizontal and near vertical rays have access to nearby transmitting regions and, ultimately, spend most of their time in $\beta_6$ or, for the infinite horizon cell, $\beta_6$ and $\beta_{19}$. 
A deeper understanding on these hierarchy shown in Fig.~\ref{fig:vbic} deserves further investigation. 

\section{Conclusions}
\label{sec:concl}

We have analytically derived a map for the scattering of incoming trajectories to an open polygonal billiard with opening angle $\alpha=\pi/2$ with finite and infinite horizon. The scattering map is defined on a partition of the coordinate space that we have obtained from the classification of the singular directions through an interval exchange transformation. Each partition is defined by a given family of trajectories sharing the same symbolic code of collisions with the edges of the billiard cell.

The scattering map constitutes an improvement for the evaluation of the billiard flow. As an example of this, we have used the map to obtain analytical expressions for the dwell time of the scattered trajectories. Furthermore, we have used the dwell time to obtain an analytical expression for the expected speed of propagation of the sub-leading ballistic front that appears in the statistics of the particle displacement. We have also looked at the emergence of these ballistic fronts and found that the sub-leading front appears as a result of incoming trajectories in regions of the partition $ \beta_{3}, \beta_{4}, \beta_6, \beta_7$ and $\beta_{12}$ for the cell with finite horizon, and $\beta_{4}, \beta_6, \beta_{12}$ and $\beta_{19}$ for the cell with infinite horizon, and a well defined symbolic hierarchy resulting from a geometric selection rule.

Our findings pave the road for a better understanding of the wild dependence that particle transport has on the geometry of the billiard, in particular of the opening angle $\alpha$.

\ack JO acknowledges partial financial support from the Australian Government Research Training Program Scholarship and from ARC Discovery Project grant DP180101512. CMM acknowledges financial support from the Spanish Government grant PID2021-127795NB-I00 (MCIU/AEI/FEDER, UE). LR gratefully acknowledges support from the Italian Ministry of University and Research (MUR) through the grant PRIN2022-PNRR project (No. P2022Z7ZAJ) “A Unitary Mathematical Framework for Modelling Muscular Dystrophies” (CUP: E53D23018070001).

\section*{References}

\bibliographystyle{iopart-num}
\bibliography{polygon}

\appendix

\newpage
\section{Partition of the domain of the scattering map}
\label{app:A}

In this appendix we give explicit definitions of the partition of $\B^+$, corresponding to incoming rays flowing in the cell through the left opening $\lb$.

\subsection{Finite horizon cell $d=1/2$}

We denoted the partition $\beta_j = \mathcal{D}_j \times \mathcal{A}_j^+$, satisfying $\B^+ = \bigcup \beta_j$, which is composed of 19 regions. Of these, the regions $\{ \beta_0,\beta_4,\beta_5,\beta_6,\beta_7,\beta_8,\beta_{13},\beta_{14},\beta_{15},\beta_{16},\beta_{17} \}$ are defined as
\begin{equation*}
  \beta_j = \mathcal{D} \times \{ \theta : h \in \mathcal{D} \ , \ \xi_j(h) < \theta < \xi_{j+1}(h) \} \ ,
\end{equation*}
where the angles $\xi_j$ are arc tangents parameterising the singular directions in each case. The remaining regions, $\{\beta_1,\beta_2,\beta_3,\beta_9,\beta_{10},\beta_{11},\beta_{12},\beta_{18}\}$, are either
\begin{enumerate}
\item[$i$)] bounded by more than two curves,
\item[$ii$)] defined on a domain $\mathcal{D}_j \subset \mathcal{D}$, or
\item[$iii$)] endlessly subdivided into regions corresponding to bouncing trajectories.
\end{enumerate}

Each of these three scenarios correspond to the occurrence of nested singular trajectories and are handled by an appropriate truncation of the region. This has been  taken into account in Sec.~\ref{sec:g0} by defining the intervals $\mathcal{D}_1 = \left(\frac{1}{6}, d\right)$, $\mathcal{D}_2 = \left(\frac{1}{8}, d\right)$, $\mathcal{D}_3 = \mathcal{D}_{10} = \left(0, \frac{1}{3}\right)$ and $\mathcal{D}_{11} = \left(0, \frac{1}{4}\right)$. In addition we define the following regions: $Y_1 = \mathcal{D} \times \{ \theta : h \in \mathcal{D} \ , \ -\Tinv{\frac{h}{d}} < \theta < -\Tinv{\frac{h + d - \delta}{d + \delta}} \}$ and $Y_7 = \mathcal{D} \times \{ \theta : h \in \mathcal{D} \ , \ \Tinv{\frac{\delta - h}{d+\delta}} < \theta < \Tinv{\frac{d - h}{d}} \}$. 

Using this, all regions appearing in Fig.~\ref{fig:chisFH} are explicitly given by
\begin{eqnarray*}
\fl  \beta_0    &=\mathcal{D} \times          \left\lbrace \theta : h \in \mathcal{D}   \ , \ -\frac{\pi}{2}         < \theta < -\Tinv{\frac{h}{d}} \right\rbrace   \\
\fl    \beta_1    &=\mathcal{D}_1 \times	     \left\lbrace \theta : h \in \mathcal{D}_1 \ , \ -\Tinv{\frac{h}{d}}    < \theta < -\Tinv{\frac{h+d}{4d}} \right\rbrace  \\
\fl    \beta_2    &=\left[\mathcal{D}_2 \times  \left\lbrace \theta : h \in \mathcal{D}_2 \ , \ -\Tinv{\frac{h+d}{4d}} < \theta < -\Tinv{\frac{h+d}{5d}} \right\rbrace \right] \cap Y_1 \\
 \fl   \beta_3    &=\left[\mathcal{D}_3 \times  \left\lbrace \theta : h \in \mathcal{D}_3 \ , \ -\Tinv{\frac{h+d}{5d}} < \theta < -\Tinv{\frac{h+d - \delta}{d + \delta}} \right\rbrace \right] \cap Y_1 \\
 \fl   \beta_4    &=\mathcal{D} \times          \left\lbrace \theta : h \in \mathcal{D}   \ , \ -\Tinv{\frac{h+d - \delta}{d + \delta}} < \theta < -\Tinv{\frac{h}{d+1}} \right\rbrace  \\
\fl    \beta_5    &=\mathcal{D} \times          \left\lbrace \theta : h \in \mathcal{D}   \ , \ -\Tinv{\frac{h}{d+1}}     < \theta < -\Tinv{\frac{h}{2d+1}} \right\rbrace  \\
 \fl   \beta_6    &=\mathcal{D} \times          \left\lbrace \theta : h \in \mathcal{D}   \ , \ -\Tinv{\frac{h}{2d+1}}    < \theta < \Tinv{\frac{d - h}{2d+1}} \right\rbrace  \\
 \fl   \beta_7    &=\mathcal{D} \times          \left\lbrace \theta : h \in \mathcal{D}   \ , \ \Tinv{\frac{d - h}{2d+1}} < \theta < \Tinv{\frac{\delta - h}{2d+\delta}} \right\rbrace  \\
\fl    \beta_8    &=\mathcal{D} \times          \left\lbrace \theta : h \in \mathcal{D}   \ , \ \Tinv{\frac{\delta - h}{2d+\delta}} < \theta < \Tinv{\frac{\delta - h}{d+\delta}} \right\rbrace \\
 \fl   \beta_9    &=\left[\mathcal{D} \times    \left\lbrace \theta : h \in \mathcal{D}   \ , \ \Tinv{\frac{\delta - h}{d+\delta}}  < \theta < \Tinv{\frac{1 - h}{2d + 1}} \right\rbrace  \right] \cap Y_7\\
 \fl   \beta_{10} &=\left[\mathcal{D}_{10} \times \left\lbrace \theta : h \in \mathcal{D}_{10} \ , \ \Tinv{\frac{1 - h}{2d + 1}}    < \theta < \Tinv{\frac{1-h}{d+1}}       \right\rbrace  \right] \cap Y_7\\
 \fl   \beta_{11} &=\mathcal{D}_{11} \times     \left\lbrace \theta : h \in \mathcal{D}_{11}   \ , \ \Tinv{\frac{1-h}{d+1}}         < \theta < \Tinv{\frac{d-h}{d}} \right\rbrace  \\ 
 \fl   \beta_{12} &=\mathcal{D} \times          \left\lbrace \theta : h \in \mathcal{D}   \ , \ \Tinv{\frac{d-h}{d}}	          < \theta < \Tinv{\frac{d + \delta - h}{d + \delta}} \right\rbrace  \\ 
 \fl   \beta_{13} &=\mathcal{D} \times          \left\lbrace \theta : h \in \mathcal{D}   \ , \ \Tinv{\frac{d + \delta - h}{d + \delta}} < \theta < \Tinv{\frac{d+1-h}{d+1}} \right\rbrace  \\ 
 \fl   \beta_{14} &=\mathcal{D} \times          \left\lbrace \theta : h \in \mathcal{D}   \ , \ \Tinv{\frac{d+1-h}{d+1}} < \theta < \Tinv{d+1-h} \right\rbrace  \\ 
 \fl   \beta_{15} &=\mathcal{D} \times          \left\lbrace \theta : h \in \mathcal{D}   \ , \ \Tinv{d+1-h}	       < \theta < \Tinv{\frac{d+\delta - h}{\delta}} \right\rbrace  \\ 
 \fl   \beta_{16} &=\mathcal{D} \times          \left\lbrace \theta : h \in \mathcal{D}   \ , \ \Tinv{\frac{d+\delta - h}{\delta}}    < \theta < \Tinv{2d+1-h} \right\rbrace  \\ 
 \fl   \beta_{17} &=\mathcal{D} \times          \left\lbrace \theta : h \in \mathcal{D}   \ , \ \Tinv{2d+1-h}			    < \theta < \Tinv{\frac{\delta + 2d - h}{\delta}} \right\rbrace  \\ 
 \fl   \beta_{18} &=\mathcal{D} \times          \left\lbrace \theta : h \in \mathcal{D}   \ , \ \Tinv{\frac{\delta + 2d - h}{\delta}} < \theta < \frac{\pi}{2} \right\rbrace 
 \end{eqnarray*}
 where for convenience, we defined $\delta = 1/2$.

In addition, the regions $\beta_{1}$, $\beta_{12}$ and $\beta_{18}$ contain a nested family of bouncing trajectories. To take this into account  the $n$-th sub-region with $n \ge 0$ is defined as
\begin{eqnarray*}
\fl     \beta_{1}^{(n)} &=\left[\mathcal{D} \times \left\lbrace \theta : h \in \mathcal{D}  \ , \ -\Tinv{\frac{h+(n+1)d}{1+(n+1)d}}     < \theta < -\Tinv{\frac{h+nd}{1+nd}}       \right\rbrace \right] \cap \beta_1 \\
\fl     \beta_{12}^{(n)} &=\left[\mathcal{D} \times \left\lbrace \theta : h \in \mathcal{D} \ , \ -\Tinv{\frac{1+d(n-1)-h}{1+nd}}	     < \theta < -\Tinv{\frac{1+nd-h}{1+(n+1)d}} \right\rbrace \right] \cap \beta_{12} \\
\fl     \beta_{18}^{(1)} &=\mathcal{D} \times \left\lbrace \theta : h \in \mathcal{D} \ , \  \Tinv{\frac{\delta + 2d - h}{\delta}} < \theta < \Tinv{\frac{1 + 2d - h}{d}}     \right\rbrace \\
\fl     \beta_{18}^{(2)} &=\mathcal{D} \times \left\lbrace \theta : h \in \mathcal{D} \ ,  \ \Tinv{\frac{1 + 2d - h}{d}}	     < \theta < \frac{\pi}{2}			\right\rbrace
\end{eqnarray*}
 All regions have been defined for an arbitrary value of $d$.

\subsection{Infinite horizon cell $d=1$}

For the cell with infinite horizon the partition consist of 16 regions. With the exception of the region $\beta_{19}$ that exists only in the case of infinite horizon, each of the remaining 15 regions have one to one correspondence with regions in the partition $\{\beta_m\}$ of Sec.~\ref{sec:finite}

As discussed in Sec.~\ref{sec:infinite}, the order of singular directions changes with respect to cell width when moving from $d = 1/2$ to $d = 1$. This resulted in some regions of the partition $\{\beta_m\}$ not being present in the infinite horizon case. To take this into account we defined a permutation vector $J(k) = 1,3,4,6,7,5,8$, for $k = 0,1,\ldots,6$. In addition, we introduce the vector $\{ \mu_{k} \}_{k=0}^{6} = \{ 1, \frac{2}{3},\frac{3}{5},\frac{1}{2},\frac{1}{2},\frac{2}{5},\frac{1}{3} \}$ and the height intervals $\mathcal{D}_{J(k)} = (0, \mu_{k})$ so that the permuted regions may be given explicitly as,
\begin{eqnarray*}
\    \Z^+_{J(0)} &= \left[ \mathcal{D}_{J(0)} \times \left\lbrace \theta : h \in \mathcal{D}_{J(0)} \  , \ -\Tinv{\frac{h}{d}}               < \theta < -\Tinv{\frac{h}{d+1}}            \right\rbrace \right] \cap A \\ 
    \Z^+_{J(1)} &= \left[ \mathcal{D}_{J(1)} \times \left\lbrace \theta : h \in \mathcal{D}_{J(1)} \  , \ -\Tinv{\frac{h}{d+1}} \}          < \theta < -\Tinv{\frac{h}{2d+1}}           \right\rbrace \right] \cap A \\ 
    \Z^+_{J(2)} &= \left[ \mathcal{D}_{J(2)} \times \left\lbrace \theta : h \in \mathcal{D}_{J(2)} \  , \ -\Tinv{\frac{h}{2d+1}}            < \theta < \Tinv{\frac{\delta-h}{2d+\delta}}\right\rbrace \right] \cap A \\ 
    \Z^+_{J(3)} &= \left[ \mathcal{D}_{J(3)} \times \left\lbrace \theta : h \in \mathcal{D}_{J(3)} \  , \ \Tinv{\frac{\delta-h}{2d+\delta}} < \theta < \Tinv{\frac{\delta-h}{d+\delta}} \right\rbrace \right] \cap A \\ 
    \Z^+_{J(4)} &= \left[ \mathcal{D}_{J(4)} \times \left\lbrace \theta : h \in \mathcal{D}_{J(4)} \  , \ \Tinv{\frac{\delta-h}{d+\delta}}  < \theta < \Tinv{\frac{d-h}{2d+1}}          \right\rbrace \right] \cap A \\ 
    \Z^+_{J(5)} &= \left[ \mathcal{D}_{J(5)} \times \left\lbrace \theta : h \in \mathcal{D}_{J(5)} \  , \ \Tinv{\frac{d-h}{2d+1}}           < \theta < \Tinv{\frac{d-h}{d+1}}           \right\rbrace \right] \cap A \\ 
    \Z^+_{J(6)} &= \left[ \mathcal{D}_{J(6)} \times \left\lbrace \theta : h \in \mathcal{D}_{J(6)} \ ,  \ \Tinv{\frac{d-h}{d+1}}            < \theta < \Tinv{\frac{\delta-h}{\delta}}   \right\rbrace \right] \cap A \\ 
\end{eqnarray*}
where $A = \mathcal{D} \times \{ \theta : h \in \mathcal{D} \ \mathrm{and} \ -\Tinv{h} < \theta < \Tinv{1 - 2h} \}$.

The regions $\beta_{0}, \beta_{12}, \beta_{13}, \beta_{14}, \beta_{15}, \beta_{16}, \beta_{17}, \beta_{18}$ coincide with their definitions for the finite cell and the region $\beta_{19}$ is defined as
\begin{equation*}
  \beta_{19} = \mathcal{D} \times \left\{ \theta : h \in \mathcal{D} \ , \ \Tinv{\frac{\delta-h}{\delta}}  < \theta < \Tinv{d-h}  \right\}
\end{equation*}
All regions have been defined for an arbitrary value of $d$.
\end{document}